\documentclass[aps,prb,times,twocolumn,amsmath,amssymb,superscriptaddress,floatfix,showpacs]{revtex4}

\usepackage{color}

\usepackage[dvips]{graphicx}
\usepackage{subfigure}
\usepackage{bbold}
\usepackage{verbatim}
\usepackage{float}
\usepackage{enumerate}
\usepackage{amsfonts}
\usepackage{multirow}

\newcommand{\dg}{^\dagger}

\begin{document}


\title{
Waves in the unseen : theory of spin excitations 
in a quantum spin-nematic 
}


\author{Andrew Smerald}

\affiliation{Okinawa Institute of Science and Technology, Onna-son, Okinawa 904-0412, Japan}
\affiliation{H.\ H.\ Wills Physics Laboratory, University of Bristol,  Tyndall Av, BS8--1TL, UK.}

\author{Nic Shannon}
\affiliation{Okinawa Institute of Science and Technology, Onna-son, Okinawa 904-0412, Japan}
\affiliation{H.\ H.\ Wills Physics Laboratory, University of Bristol,  Tyndall Av, BS8--1TL, UK.}


\date{\today}


\begin{abstract}  
The idea that a quantum magnet could act like a liquid crystal, breaking
spin-rotation symmetry without breaking time-reversal symmetry, holds
an abiding fascination.
However, the very fact that spin nematic states do not break 
time-reversal symmetry renders them ``invisible'' to the most common
probes of magnetism --- they do {\it not} exhibit magnetic Bragg peaks, 
a static splitting of lines in NMR spectra, or oscillations in $\mu$SR.
Nonetheless, as a consequence of breaking spin-rotation symmetry,
spin-nematic states {\it do} possess a characteristic spectrum of dispersing
excitations which {\it could} be observed in experiment.
With this in mind, we develop a symmetry-based description of long-wavelength 
excitations in a spin-nematic state,  based on an {\sf SU(3)} generalisation 
of the quantum non-linear sigma model.
We use this field theory to make explicit predictions for inelastic neutron scattering, 
and argue that the wave-like excitations it predicts could be used to identify the 
symmetries broken by the otherwise unseen spin-nematic order.
\end{abstract}


\maketitle


\section{Introduction}


The search for quantum spin liquids, magnets which do not order at {\it any} temperature, 
has become one of the cause c\'el\`ebre of modern physics\cite{lee08}.
Another equally intriguing  
possibility is that the spins  
of a quantum 
magnet {\it do} order, but in a way which does not transform like a spin.
Such a state would be almost invisible to the usual probes
of magnetism, and could therefore appear as a ``hidden order''.
A concrete example of this is the quantum spin-nematic --- a magnetic analogue of 
a liquid crystal\cite{blume69,chen71,andreev84,papanicolau88,chandra91,barzykin91, barzykin93}.

\begin{figure}[h]
\centering
\includegraphics[width=0.45\textwidth]{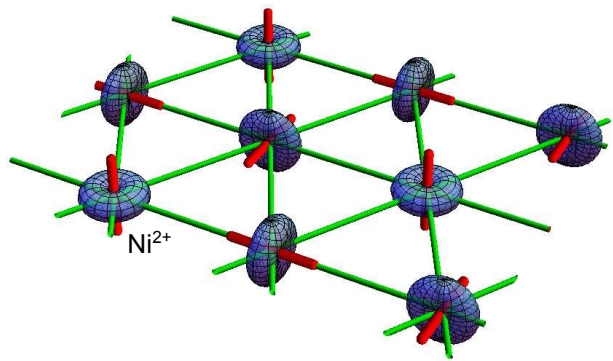}
\caption{\footnotesize{(Color online). 
3-sublattice antiferroquadrupolar (AFQ) spin-nematic state, 
found in the spin-1 bilinear-biquadratic model on a triangular lattice, 
and studied in the context of NiGa$_2$As$_2$ [\onlinecite{tsunetsugu06,tsunetsugu07,lauchli06}].  
The probability distribution of spin fluctuations (shown as a blue surface) 
define orthogonal directions on neighbouring sites.  
The directors describing this AFQ state are represented by red cylinders.
}}
\label{fig:NiGa2As2}
\end{figure}
\begin{figure}[h]
\centering
\includegraphics[width=0.45\textwidth]{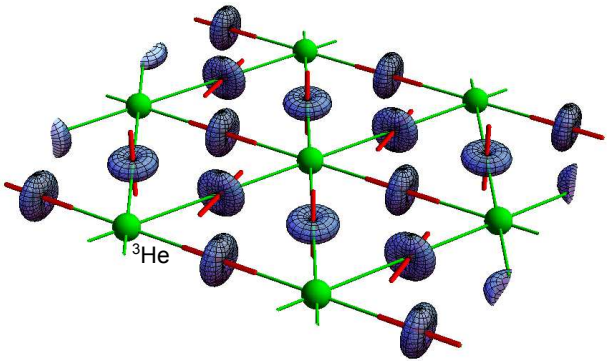}
\caption{\footnotesize{(Color online). 
The 3-sublattice, bond-centred, antiferroquadrupolar
 (AFQ) spin-nematic state proposed to exist in 
thin films of $^3$He [\onlinecite{momoi06,momoi12}]. 
This system can be modelled using a multiple spin exchange model 
of $^3$He atoms with nuclear \mbox{\mbox{spin-1}/2} (represented 
by green spheres), on a two-dimensional triangular lattice.   
For a range of parameters bordering on ferromagnetism, 
the ground state of this model is a 3-sublattice AFQ order in which 
spin fluctuations (shown as a blue surface) are orthogonal on 
neighbouring bonds.
The directors describing this AFQ state are represented by red cylinders.
}}
\label{fig: 3He}
\end{figure}
\begin{figure}[ht]
\centering
\includegraphics[width=0.48\textwidth]{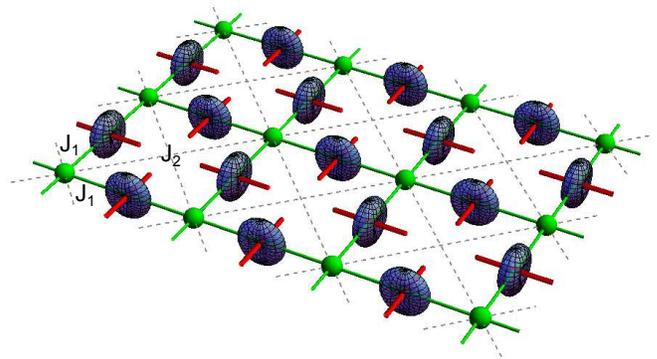}
\caption{\footnotesize{(Color online). 
Two-sublattice, bond-centred, antiferroquadrupolar
 (AFQ) spin-nematic state found 
bordering the ferromagnetic state in both the \mbox{\mbox{\mbox{spin-1}/2}} 
\mbox{$J_1-J_2$} Heisenberg model and the \mbox{\mbox{spin-1}/2}
 multiple spin exchange model on the square lattice
\cite{shannon06,shindou09,ueda09,shindou11,shindou13}. 
Magnetic ions are denoted by green spheres, and the probability distribution 
of spin fluctuations on each bond is shown as a blue surface.  
The directors describing this AFQ state are represented by red cylinders.
}}
\label{fig:J1J2nematic}
\end{figure}

Conventional nematic order is associated with the directional 
order of rod- or disk-like molecules.
Spin-nematic order occurs where the {\it fluctuations} of a spin mimic a uniaxial molecule, selecting an axis without selecting a direction along it.
For example, a system could exhibit fluctuations such that 
\mbox{$\langle ({\sf S}^{\sf x})^2 \rangle=\langle ({\sf S}^{\sf y})^2  \rangle \neq \langle ({\sf S}^{\sf z})^2 \rangle$} 
while maintaining \mbox{$\langle {\bf {\sf S}} \rangle=0$}.
Such a phase would break spin-rotation symmetry {\it without} 
breaking time-reversal symmetry.
This particular type of spin-nematic state can be described as ``ferro-quadrupolar'' (FQ), 
since the fluctuations form a quadrupole moment of ${\bf {\sf S}}$ with a common
axis on all sites (for an introduction, see [\onlinecite{penc11}]).
More generally, quadrupole moments tend to select orthogonal axes.
Examples of this kind of ``antiferroquadrupolar'' (AFQ) order are shown in 
Fig.~\ref{fig:NiGa2As2}--\ref{fig:J1J2nematic}.


There are now good theoretical reasons to believe that spin-nematic order should 
occur in a range of low-dimensional and frustrated systems.
However, because the spin-nematic state does not break time-reversal symmetry, it is {\it ``invisible''} to the tests 
commonly used to discern magnetic order, namely the existence of magnetic Bragg peaks in 
elastic neutron scattering, the splitting of lines in NMR spectra, or through the asymmetry of 
oscillations in $\mu$SR spectra.
Nevertheless, since excitations of the spin-nematic state induce a fluctuating dipole moment, spin-nematic
order can, in principle, be detected by {\it dynamic} probes of magnetism, such as inelastic neutron 
scattering or the NMR $1/T_1$ relaxation rate.
This hints at an interesting question --- if we can't measure the symmetry breaking in a spin-nematic
state directly, can we infer it from the associated excitations ?
%


In this paper, we set aside all questions of the microscopic origin of spin-nematic order, and 
attempt to say something about what the excitations of a spin-nematic state would look 
like, assuming it existed.
To this end we develop a phenomenological, symmetry-based description of long-wavelength excitations 
in AFQ spin-nematic states, based on an {\sf SU(3)} generalisation of the quantum non-linear sigma model, 
and use it to make concrete predictions for inelastic neutron scattering and the dynamical quadrupolar susceptibility.


We build on a long history of studying spin-nematic states.
In one dimension, theoretical studies support the existence of Luttinger liquids with dominant 
power-law correlations of spin-quadrupole moments (and in some cases, higher-order spin-multipoles), 
in frustrated ferromagnetic spin
chains\cite{chubukov87,chubukov90, chubukov91-JPCM, chubukov91-PRB, cabra00, 
heidrich-meisner06, lu06, kuzian07, vekua07, kecke07, hikihara08, sudan09, 
syromyatnikov12, nishimoto-arXiv}, 
in \mbox{spin-1/2} ladders with cyclic exchange~\cite{nersesyan98,totsuka12} 
and for \mbox{spin-1} models with biquadratic interactions\cite{fath98,lauchli06-PRB,manmana11}.


In two dimensions, theoretical studies suggest the existence of a bond-centred, 
spin-nematic ground state in models of \mbox{\mbox{spin-1}/2} frustrated ferromagnets on the 
square\cite{shannon06,shindou09,ueda09,shindou11,shindou13} [Fig.~\ref{fig:J1J2nematic})]
and the triangular lattices\cite{momoi06,momoi12}, and of a generalised chiral nematic phase 
on the square lattice~\cite{chandra91,laeuchli05}.   
Similarly, two-dimensional, \mbox{spin-1} models with biquadratic interactions support $T=0$ nematic 
order\cite{papanicolau88,onufrieva85,joshi99,harada02,tsunetsugu06,tsunetsugu07,lauchli06,toth10,toth12,kaul12}.
Entropy-driven nematic order has also been widely studied in the context of the classical Heisenberg model 
on the Kagome lattice~\cite{chalker92,zhitomirsky08}.  


In three dimensions, quantum Monte Carlo calculations find evidence for a \mbox{spin-1} nematic state in the 
bilinear-biquadratic model\cite{harada02}, and classical spin-nematic states have been proposed 
on various frustrated lattices\cite{chalker92,moessner98,moessner982,hopkinson07,zhitomirsky08,shannon10}.
Weakly-coupled chains in magnetic field also exhibit long-range spin-nematic order\cite{sato13}.


Recently, the study of spin-nematic order has been re-energised 
by the proposal that it might occur in a number of real materials.
The unusual magnetic ground state of the \mbox{spin-1} layered magnetic 
insulator NiGa$_2$S$_4$~\cite{nakatsuji05} has been discussed in terms of 
both FQ\cite{bhattacharjee06,lauchli06} and AFQ\cite{tsunetsugu06,tsunetsugu07} 
order [cf. Fig.~\ref{fig:NiGa2As2}], and spin-freezing in the presence of FQ correlations\cite{stoudenmire09}, 
with the bilinear-biquadratic model on a triangular lattice used as a prototype for calculations.
Exact diagonalization studies of the relevant multiple spin-exchange model
suggest that the ``spin liquid'' ground state of thin films of $^3$He might be associated with 
a 3-sublattice, bond-centred, AFQ phase\cite{momoi06,momoi12}, [cf. Fig.~\ref{fig: 3He})].
Related calculations suggest that a 2-sublattice, bond-centered, AFQ spin-nematic state 
[cf. Fig.~\ref{fig:J1J2nematic}] might also be also be realised in the spin-1/2 frustrated 
Heisenberg model relevant to a family of square lattice vanadates~\cite{shannon06}.
And finally, magnetisation measurements on the spin-chain system LiCuVO$_4$ show a 
phase transition close to saturation, which has been interpreted as the onset of a 
bond-centred, AFQ state\cite{svistov10,zhitomirsky10}.


In parallel with this new work on magnetic insulators, there has been an explosion of interest in 
electronic-nematic states in itinerant transition-metal compounds, and a resurgence 
of interest in the study of multipolar ``hidden order'' phases in rare-earth materials\cite{santini09}.
Since these systems are typically metallic and/or subject to strong spin-orbit coupling, 
somewhat different considerations apply, and we will not attempt to review either 
subject here.
We concentrate instead on local moments with a high degree of spin-rotational symmetry.

%
\begin{figure*}[ht]
\includegraphics[width=0.8\textwidth]{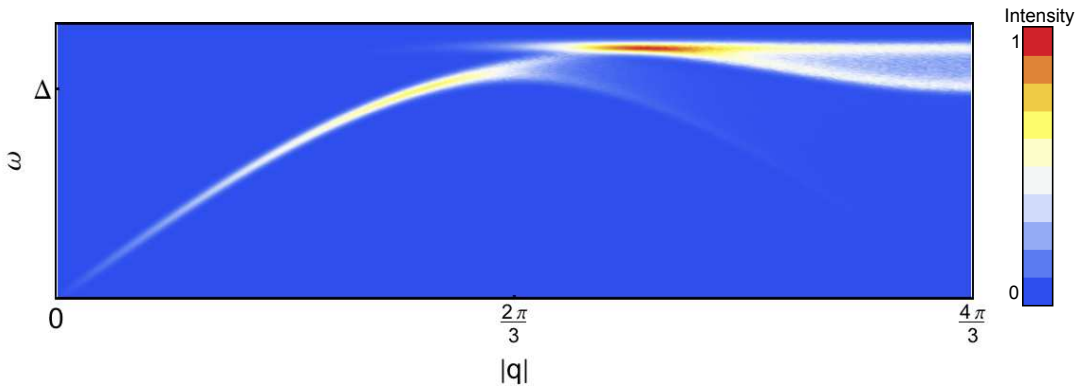}
\caption{\footnotesize{(Color online). 
Prediction for inelastic neutron-scattering from a powder sample of a triangular-lattice 
magnet with a 3-sublattice antiferroquadrupolar (AFQ) spin-nematic ground state. 
Spectral weight is found predominantly in the upper, spin-wave band, but vanishes approaching ${\bf q}=0$.
Intensity in the lower, quadrupole-wave band is weaker, 
and vanishes approaching the magnetic ordering vector $|{\bf q}|=4\pi/3$.
Results are taken from the linear ``flavour wave'' analysis of the \mbox{spin-1} bilinear-biquadratic 
model [Eq.~(\ref{eq:H-BBQ-triangle})], as described in Ref.~[\onlinecite{tsunetsugu06}] and 
Section~\ref{sec:3-sublattice-AFQ} of this paper, for parameters $J_1=1$, $J_2=1.22$. 
The prediction for the dynamic structure factor $S({\bf q}, \omega)$ has been 
integrated over angle and convoluted with a Gaussian of FWHM 
\mbox{$ \omega=0.042 \Delta$}, where $\Delta = 6\sqrt{J_2(J_1-J_2)}$.  
}}
\label{fig:powder-3-sublattice}
\end{figure*}
%


While this brings some simplifications, the microscopic models needed to describe thin films 
of $^3$He~[\onlinecite{momoi12}] and LiCuVO$_4$~[\onlinecite{zhitomirsky10, syromyatnikov12}] are 
already very complex, with dominant nearest-neighbour ferromagnetic interactions frustrated by a large 
number of competing antiferromagnetic exchange pathways.
The complexity of these models points to the need for a phenomenological description 
of AFQ order which makes explicit the physical nature of its excitations, and parameterizes them 
in terms of the smallest possible number of experimentally-measurable parameters.


In this article, we develop a symmetry-based description of the long-wavelength 
excitations of 3-sublattice AFQ order on the triangular lattice.
Our approach, based on an {\sf SU(3)} generalisation of the quantum non-linear 
sigma model, could be applied equally to the spin-1 magnet
NiGaS$_2$~[\onlinecite{lauchli06,tsunetsugu06,tsunetsugu07}], or to 
thin films of $^3$He~[\onlinecite{momoi06,momoi12}].  
With minor modifications, the action we derive also offers a description of the 
2-sublattice AFQ order proposed to occur in LiCuVO$_4$~[\onlinecite{zhitomirsky10, syromyatnikov12}], 
and square lattice frustrated ferromagnets~\cite{shannon06}.
In fact, it can be modified to describe any system where spin-quadrupoles 
display short- or long-range, non-collinear order. 
The only requirement is that the Hamiltonian either has a continuous 
symmetry (e.g. SU(2) or U(1)), or is close to having a continuous 
symmetry. 


In order to demonstrate the validity of this approach, we show explicitly how our sigma-model like
action can be derived from a microscopic model exhibiting 3-sublattice AFQ order, the \mbox{spin-1} 
bilinear-biquadratic (BBQ) model on a triangular lattice~\cite{lauchli06,tsunetsugu06,tsunetsugu07}. 
At long wavelength, the resulting continuum theory exactly reproduces published results 
for ``flavour-wave'' analysis of the lattice model~\cite{tsunetsugu07,lauchli06}.  
However, the continuum theory is both independent of the ``flavour-wave'' theory and far more 
general, and could equally well be parametrised from experiment, or from analysis of a 
more-complicated microscopic model where the ``flavour-wave'' approach is not applicable.


Good reviews exist of ``flavour wave'' techniques for spin-nematic 
order~\cite{penc11}, but sigma-model approaches have yet to be reviewed,
and have so far been restricted to FQ order~\cite{ivanov03,ivanov07,ivanov08,baryakhtar13}.  
We therefore provide a complete and pedagogical account of the 
steps needed to derive a non-linear sigma-model description of AFQ order.


The fact that different branches of excitation correspond to different rotations of the order 
parameter, allows us to assign each branch of excitations a clear physical meaning.
In the case of 3-sublattice AFQ order, we identify two, physically-distinct types of magnetic excitation --- 
three degenerate branches of ``quadrupole waves'', the gapless, linearly-dispersing Goldstone modes of 
AFQ order, and three degenerate branches of gapped, high-energy ``spin-wave'' excitations.
The spin-wave excitations have a substantial fluctuating dipole moment, and so should be 
clearly visible in experiment.


Having constructed a general theory for the long-wavelength excitations of the 3-sublattice  
AFQ spin-nematic states, we are in position to make explicit predictions for inelastic neutron 
scattering experiments.
An example is given in Fig.~\ref{fig:powder-3-sublattice}.
Observation of these features in experiment would provide strong evidence for spin-nematic order, 
and a means of distinguishing between different types of spin-nematic states.


We also show predictions for the dynamic quadrupole susceptibility.
This may be measurable using, for example, resonant x-ray scattering.
%


When calculating the experimental response we neglect interaction between the modes.
Since we are primarily interested in the universal, long wavelength features it is expected that this is a good approximation.
We will return to the role of interactions in a future publication\cite{smerald-unpub}.
We note that any treatment of the 2-particle continuum excitations {\it must} take the role of 3- and 4-particle interactions into account if it is to obey the symmetry-constrained sum rules, and for this reason we do not discuss the continuum in this publication.
%

  
The remainder of this article is structured as follows.  
In Section~\ref{sec:3-sublattice-AFQ} we develop a theory of long-wavelength 
excitations in a 3-sublattice AFQ spin-nematic state.
In Section~\ref{sec:neutrons}, we explore how the excitations of each of
these states would manifest themselves in inelastic neutron scattering experiments. 
In Section~\ref{sec:quad-suscep} we consider the dynamical quadrupolar susceptibility. 
Finally, in Section~\ref{sec:conclusions} we conclude with a summary of results and 
discussion of their experimental context.
Readers who are already expert in sigma models, or simply uninterested in 
these technical details, are invited to pass directly to Section~\ref{sec:neutrons}, 
where all key results are summarised.
Results for spin-nematic states in 2-sublattice states, in applied magnetic field and predictions for the NMR $1/T_1$ relaxation 
rate, will be presented in a separate publication\cite{smerald-unpub}.


\section{Continuum theory of 3-sublattice AFQ order}
\label{sec:3-sublattice-AFQ}


\subsection{Minimal microscopic model}
\label{sec:BBQ-triangle}


To keep our continuum theory grounded in microscopic reality, it is helpful to be able 
to derive it directly from a concrete lattice model, even though the resulting 
field theory will have far broader applicability.
The simplest microscopic model with an AFQ ground state is the 
\mbox{spin-1} bilinear-biquadratic (BBQ) model on a triangular 
lattice\cite{blume69,papanicolau88,lauchli06}.  
This model is defined by
\begin{align}
\mathcal{H}^{\sf BBQ}_\triangle 
   = \sum_{\langle ij\rangle} J_1 {\bf S}_i.{\bf S}_j +J_2 \left( {\bf S}_i.{\bf S}_j \right)^2,
\label{eq:H-BBQ-triangle}
\end{align}
where the sum on $\langle ij\rangle$ runs over the nearest-neighour bonds of a triangular lattice.

\begin{figure}[t]
\centering
\includegraphics[width=0.48\textwidth]{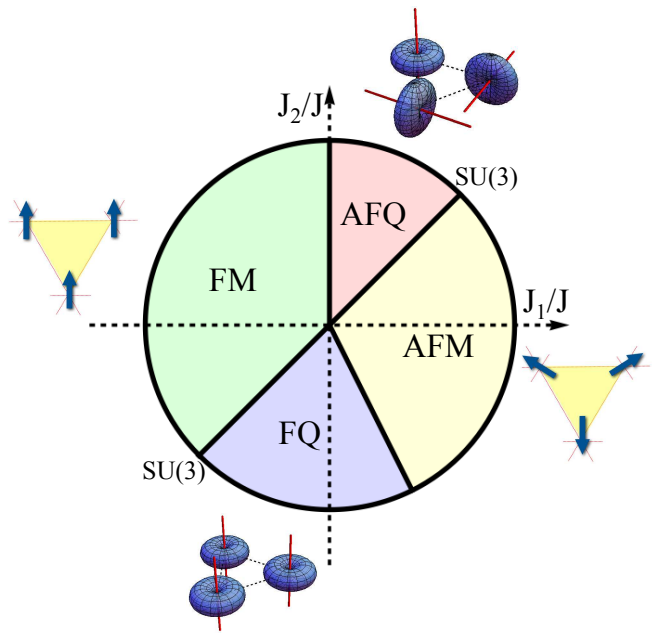}
\caption{\footnotesize{(Color online). 
The mean-field, ground-state phase diagram for the \mbox{spin-1} bilinear-biquadratic (BBQ)
model on a triangular lattice ${\mathcal H}_\triangle^{\sf BBQ}$ [Eq.~(\ref{eq:H-BBQ-triangle})], 
following Ref~[\onlinecite{penc11,lauchli06}], showing two distinct regions of spin-nematic order. 
In the ferro-quadrupolar (FQ) 
phase, all directors are aligned.
In the 3-sublattice antiferroquadrupolar (AFQ) phase, 
directors on different sublattices are orthogonal. 
The model also supports two conventional magnetic phases --- the ferromagnet (FM); and 
the 3-sublattice ``120$^\circ$'' antiferromagnet (AFM).  
For $J_1=J_2$, 
the symmetry of the model is increased 
from {\sf SU(2)} to {\sf SU(3)}.
}}
\label{fig:phasediag}
\end{figure}

The  mean-field phase diagram for the spin-1 BBQ model on a triangular lattice\cite{penc11,lauchli06}, 
reproduced in Fig.~\ref{fig:phasediag}, exhibits an extended region of 3-sublattice AFQ order for \mbox{$J_2 > 0$}, 
terminating in a point for \mbox{$J_1 = J_2$} where the symmetry of the model is enlarged from ${\sf SU(2)}$ to 
${\sf SU(3)}$~[\onlinecite{papanicolau88}].
AFQ order is accompanied by a ferroquadrupolar (FQ) phase for \mbox{$J_2 < 0$}.
Conventional ferromagnetic (FM) and 3-sublattice ``120$^\circ$'' antiferromagnetic (AFM)
phases separate these two spin-nematic states.
A very similar phase diagram is found in exact diagonalisation\cite{lauchli06}, and the existence
of AFQ and FQ phases for closely related BBQ models has been independently confirmed by 
density matrix renormalisation group calculations\cite{bauer11}, and quantum Monte Carlo 
simulations\cite{harada02}. 


For $J_2 > J_1>0$, $\mathcal{H}^{\sf BBQ}_\triangle$ [Eq.~(\ref{eq:H-BBQ-triangle})]
favours states in which the quadrupole moments 
on neighbouring sites take on perpendicular directions.
The relative simplicity 
of this model follows from the fact that each spin-1 can form a quadrupole by itself, 
and the triangular lattice is tripartite, and so naturally supports a 3-sublattice state
in which all quadrupoles are orthogonal to one another.
The fact that an approximate ground-state wave function can be written in a site-factorized 
form\cite{penc11} makes it possible to calculate physically interesting quantities
perturbatively from the Hamiltonian using ``flavour-wave'' theory\cite{papanicolau88,lauchli06,papanicolau84,onufrieva85,joshi99,tsunetsugu06,tsunetsugu07,li07,butrim08} --- 
the ${\sf SU(3)}$ generalization of the more usual ${\sf SU(2)}$ spin-wave theory. 


The ``flavour-wave'' approach does not generalise easily to the complicated \mbox{\mbox{spin-1}/2} 
models that are relevant to systems such as $^3$He and LiCuVO$_4$.
However it provides an important benchmark for the field-theoretical approach developed
in this article.
In what follows we briefly review some of the features of the spin-1 BBQ model on a triangular lattice,
including a useful mean-field parametrisation in terms of spin coherent states, which makes explicit 
the director nature of the order parameter\cite{ivanov03,lauchli06,penc11}.


Following [\onlinecite{lauchli06,tsunetsugu06,tsunetsugu07}] the Hamiltonian, Eq.~(\ref{eq:H-BBQ-triangle}), 
can be rewritten in the form,
\begin{align}
\mathcal{H}^{\sf BBQ}_\triangle = 
   \sum_{\langle ij\rangle} \left(J_1-\frac{J_2}{2} \right) {\bf S}_i \cdot {\bf S}_j  
   + \frac{J_2}{2} {\bf Q}_i \cdot {\bf Q}_j  +\frac{J_2}{3}S^2(S+1)^2,
\label{eq:HinQS}
\end{align}
where the quadrupole operator ${\bf Q}$ is given by,
\begin{align}
{\bf Q}&=
 \left(
\begin{array}{c}
Q^{\sf x^2-y^2} \\
Q^{\sf 3z^2-r^2} \\
Q^{\sf xy} \\
Q^{\sf yz} \\
Q^{\sf xz}
\end{array}
\right)=
 \left(
\begin{array}{c}
(S^{\sf x})^2 - (S^{\sf y})^2 \\
\frac{1}{\sqrt{3}}[2(S^{\sf z})^2-(S^{\sf x})^2 - (S^{\sf y})^2] \\
S^{\sf x}S^{\sf y}+S^{\sf y}S^{\sf x} \\
S^{\sf y}S^{\sf z}+S^{\sf z}S^{\sf y} \\
S^{\sf x}S^{\sf z}+S^{\sf z}S^{\sf x}
\end{array}
\right).
\end{align}
The operator ${\bf Q}$ encodes the 5 linearly independent degrees of freedom contained in the traceless, symmetric tensor,
\begin{align}
Q^{\alpha \beta} = -\frac{2}{3}S(S+1) \delta ^{\alpha \beta} + S^\alpha S^\beta +S^\beta  S^\alpha.
\end{align}

It is common practice to parametrise the two magnetic exchange interactions as,
\begin{align}
J_1=\bar{J} \cos \theta, \quad J_2=\bar{J} \sin \theta,
\end{align}
and to plot phase diagrams on a circle, as in Fig.~\ref{fig:phasediag}.   
In this article we concentrate on the AFQ phase, bounded by the ${\sf SU(3)}$ point at $\theta=\pi/4$.


Since spin-nematic states are time-reversal invariant, it is useful to introduce a set of basis 
states that respect this symmetry. 
Following\cite{papanicolau88,ivanov03,penc11}, we consider the following linear superpositions of the 
usual \mbox{spin-1} basis states,
\begin{align}
|x\rangle=i\frac{|1\rangle-|\bar{1}\rangle}{\sqrt{2}}, \ |y\rangle=\frac{|1\rangle+|\bar{1}\rangle}{\sqrt{2}}, \ |z\rangle=-i|0\rangle.
\end{align}
A general wavefunction for an \mbox{spin-1} spin at a site $j$ can then be written in the form,
\begin{align}
|{\bf d}_j \rangle= d_j^{\sf x} |x \rangle +d_j^{\sf y} |y \rangle+d_j^{\sf z} |z \rangle,
\end{align}
where ${\bf d}_j=(d_j^{\sf x},d_j^{\sf y},d_j^{\sf z})$ is a 3 vector of complex numbers.  
It is sometimes convenient to write this out explicitly in real and 
imaginary components as,
\begin{align}
{\bf d}_j = {\bf u}_j +i{\bf v}_j.
\end{align}
Requiring the wavefunction to be normalised gives the constraint,
\begin{align}
{\bf d}_j \cdot \bar{{\bf d}}_j=1 \quad \mathrm{or} \quad {\bf u}_j^2 +{\bf v}_j^2=1,
\label{eq:lenthfix}
\end{align}
while the overall phase is set by the equation,
\begin{align}
 {\bf d}_j^2=\bar{{\bf d}}_j^2 \quad \mathrm{or} \quad {\bf u}_j \cdot {\bf v}_j=0.
 \label{eq:phasefix}
\end{align}
Since the phase does not affect any physical observables, one is free to choose this convenient value.
As a consequence of Eq.~(\ref{eq:lenthfix}) and Eq.~(\ref{eq:phasefix}), there are 4 degrees of 
freedom associated with each site.


Within the spin-coherent state framework, the operator products appearing in the Hamiltonian, 
Eq.~(\ref{eq:HinQS}), can be calculated as
\begin{align}
 {\bf S}_i.{\bf S}_j  &= |{\bf d}_i \cdot \bar{{\bf d}}_j|^2 - |{\bf d}_i\cdot{\bf d}_j|^2 \nonumber \\
 {\bf Q}_i.{\bf Q}_j  &= |{\bf d}_i \cdot \bar{{\bf d}}_j|^2+|{\bf d}_i\cdot{\bf d}_j|^2 -\frac{2}{3},
\label{eq:QS-d}
\end{align}
where the spin value has been set to \mbox{spin-1}.  
As a result, the Hamiltonian is,
\begin{align}
 \mathcal{H}^{\sf BBQ}_\triangle  
   = \sum_{\langle ij\rangle} J_1|{\bf d}_i \cdot \bar{{\bf d}}_j|^2 +(J_2-J_1)|{\bf d}_i\cdot{\bf d}_j|^2 +J_2.
\label{eq:averageH}
\end{align}
By minimising this equation, a mean-field, low temperature phase diagram can be mapped out, 
as shown in Fig.~\ref{fig:phasediag}).  


Purely real or purely imaginary values of ${\bf d}$ correspond to static nematic states, in which the 
quadrupole operators take on finite expectation values, but the spin-dipole operators do not.  
The associated director is parallel to the ``director vector'', ${\bf d}$.  
When ${\bf d}$ has both real and imaginary components, this corresponds to mixing in a non-zero, 
static dipole moment, given within the coherent state representation by,
\begin{align}
 {\bf S}_j  = 2 {\bf u}_j \times {\bf v}_j.
 \label{eq:Suv}
\end{align}
The largest dipole moment occurs when ${\bf u}$ and ${\bf v}$ are equal in magnitude (although 
even in this state there remain quadrupole operators with non-zero expectation values).


The physical observables in the system are expectation values of the dipole and quadrupole 
operators, $ {\bf S} $ and $ {\bf Q} $.  
It is useful to write these in the coherent state representation, terms of the vectors ${\bf d}$, ${\bf u}$ and ${\bf v}$, as,
\begin{align}
 &\left(
\begin{array}{c}
 S^{\sf x}  \\
 S^{\sf y} \\
 S^{\sf z} \\
 Q^{\sf x^2-y^2} \\
 Q^{\sf 3z^2-r^2} \\
 Q^{\sf xy} \\
 Q^{\sf yz} \\
 Q^{\sf xz}
\end{array}
\right)
=
 \left(
\begin{array}{c}
id^{\sf z}\bar{d}^{\sf y}-id^{\sf y}\bar{d}^{\sf z} \\
id^{\sf z}\bar{d}^{\sf x}-id^{\sf x}\bar{d}^{\sf z} \\
id^{\sf x}\bar{d}^{\sf y}-id^{\sf y}\bar{d}^{\sf x} \\
|d^{\sf y}|^2-|d^{\sf x}|^2 \\
\frac{1}{\sqrt{3}}(|d^{\sf x}|^2+|d^{\sf y}|^2 -2|d^{\sf z}|^2) \\
d^{\sf x}\bar{d}^{\sf y}+d^{\sf y}\bar{d}^{\sf x} \\
d^{\sf y}\bar{d}^{\sf z}+d^{\sf z}\bar{d}^{\sf y} \\
-d^{\sf x}\bar{d}^{\sf z}-d^{\sf z}\bar{d}^{\sf x} 
\end{array}
\right) = \nonumber \\
&
 \left(
\begin{array}{c}
 2(u^{\sf y} v^{\sf z}-v^{\sf y} u^{\sf z}) \\
 2(u^{\sf z} v^{\sf x}-v^{\sf z} u^{\sf x}) \\
 2(u^{\sf x} v^{\sf y}-v^{\sf x} u^{\sf y}) \\
(u^{\sf y})^2+(v^{\sf y})^2-(u^{\sf x})^2-(v^{\sf x})^2 \\
\frac{1}{\sqrt{3}}[ (u^{\sf x})^2+(v^{\sf x})^2+(u^{\sf y})^2+(v^{\sf y})^2-2(u^{\sf z})^2-2(v^{\sf z})^2] \\
2(u^{\sf x}u^{\sf y}+v^{\sf x}v^{\sf y}) \\
2(u^{\sf y}u^{\sf z}+v^{\sf y}v^{\sf z})  \\
-2(u^{\sf x}u^{\sf z}+v^{\sf x}v^{\sf z})  
\end{array}
\right)
\label{eq:SQduvmatrix}
\end{align}


\subsection{Continuum theory at the ${\sf SU(3)}$ point}
\label{sec:SU3}

\subsubsection{Why start here ?}


For $J_1=J_2$, the symmetry of the spin-1 BBQ model 
${\mathcal H}^{\sf BBQ}_\triangle$ [Eq.~(\ref{eq:H-BBQ-triangle})] is enlarged from 
${\sf SU(2)}$ to ${\sf SU(3)}$.
Exactly at this point, the ground states of ${\mathcal H}^{\sf BBQ}_\triangle$ include 
both the 3-sublattice AFQ state {\it and} the 3-sublattice ``120$^\circ$'' 
N\'eel antiferromagnet.
Moreover, generic 3-sublattice ground states can be constructed from both dipole and 
quadrupole moments of spins.
These physically distinct building blocks are connected by ${\sf SU(3)}$ rotations 
that transform ${\bf S}$ into ${\bf Q}$, --- and vice versa --- as well as rotating 
one spin (or quadrupole) configuration into another.
These ${\sf SU(3)}$ rotations are precisely what is needed to describe
the long-wavelength excitations of spin-nematic order, and the ${\sf SU(3)}$ 
point ($J_1=J_2$) therefore provides a very natural starting point for 
building a continuum theory of 3-sublattice AFQ order.


In the remainder of Section~\ref{sec:SU3} below, we construct 
a sigma model description of long-wavelength 
excitations of 3-sublattice AFQ order at the ${\sf SU(3)}$ point. 
We arrive at a field theory comprising of six
identical, linearly-dispersing Goldstone modes, associated with 
rotations of a triad of ${\bf d}$ vectors. 
Then, in Section~\ref{sec:SU2}, we explore the consequence of those 
terms in the Hamiltonian which break this ${\sf SU(3)}$ symmetry 
down to the more generic ${\sf SU(2)}$, introducing these as 
perturbations about the ${\sf SU(3)}$ point.
This leads to a completely general theory of long-wavelength 
excitations in a 3-sublattice AFQ state, comprising three 
gapless Goldstone modes and three gapped spin-wave excitations.


The structure of this field theory is completely determined by the symmetries
of the order parameter, and therefore independent of its derivation.
However starting from the  ${\sf SU(3)}$ point of the spin-1 BBQ model
allows us to achieve a controlled derivation of a field theory for a 3-sublattice 
AFQ state from a microscopic model, in a way which keeps the physical nature 
of its excitations in view.
This approach draws inspiration from earlier work on FQ 
order in one dimension\cite{ivanov03,ivanov07,ivanov08}, and for the 3-sublattice 
120$^\circ$ AFM state on the triangular lattice\cite{dombre89,apel92}.
In order to keep the text accessible and reasonably self-contained, the 
necessary steps are described in some detail below.


\subsubsection{Brief summary of calculation}
\label{sec:summary}

\begin{figure}[t]
\centering
\includegraphics[width=0.4\textwidth]{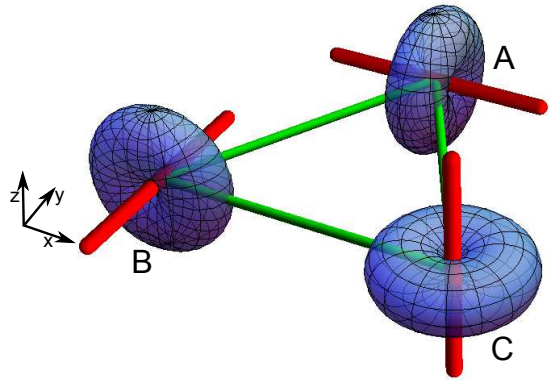}
\caption{\footnotesize{(Color online). 
The basic building-block of 3-sublattice antiferroquadrupolar (AFQ) order 
--- a triangular plaquette with directors (red cylinders) 
orientated as described in Eq.~(\ref{eq:xyzgs}). 
These directors  describe the spontaneous breaking of spin-rotation symmetry
in the AFQ phase, and are orthogonal on each of the three sublattices. 
The probability of a spin fluctuation vanishes parallel to the directors, and is maximal 
in the plane perpendicular to them --- the associated probability distribution is shown 
as a blue surface.  
}}
\label{fig:AFNtriangle}
\end{figure}

Before embarking on the calculation, it is useful to briefly summarise the main steps.
We start with a single triangular plaquette, which hosts a triad of orthogonal director vectors, 
and define matrices that describe all the physically relevant, infinitesimal rotations of this triad 
in the complex vector space of ${\bf d}$ (ie. those spanning the coset ${\sf SU(3)}/{\sf H}$, where 
${\sf H}$ defines the isotropy subgroup).
By the successive action of these matrices, {\it any} physical configuration of the three directors 
can be accessed.
Some of these matrices perform global rotations of the director triad, 
within its complex vector space, and therefore leave the energy invariant.
The remainder perform local rotations of the director configuration and thus change the energy 
of the configuration [see Fig.~\ref{fig:AFNtriangle} and Fig.~\ref{fig:AFQ-OPdynamics}].
In analogy with the collinear antiferromagnet\cite{fradkin,auerbach}, which undergoes a local 
ferromagnetic canting, these matrices can be described as a `canting' of the orthogonal director 
configuration.


The triangular plaquette acts as the basic unit from which to build the triangular lattice [see Fig.~\ref{fig:triangularlattice}].
By defining fields at the centre of plaquettes, it is possible to move from a lattice theory written in 
terms of a Hamiltonian to a continuum theory in terms of a Lagrangian.
The fields inherit the properties of the rotation matrices.
As in the case of the collinear antiferromagnet\cite{fradkin,auerbach}, in moving from the lattice Hamiltonian to the continuum Lagrangian, it is necessary to introduce a dynamical term, which arises from the quantum mechanical overlap of director configurations. 


Since we wish to describe the low temperature excitations of the antiferroquadrupolar state, it is reasonable to assume that the directors are  approximately orthogonal to one another on short lengthscales.
In consequence, the Lagrangian can be expanded in terms of the `canting' fields.
These can then be eliminated by a Gaussian integral, and the resulting action is an ${\sf SU(3)}$ symmetric non-linear sigma model.


One way to gain a better physical understanding of the resulting theory is to linearise the fields.
This allows a natural division of the modes into those with predominantly quadrupole-fluctuation character and those with spin-fluctuation character.
This forms the starting point for calculations of the experimental signatures that could prove the existence of nematic order [see Section~\ref{sec:neutrons}].


\subsubsection{Structure of the ground state manifold}
\label{sec:ground-state-manifold}


The order parameter for the AFQ phase of ${\mathcal H}^{\sf BBQ}_\triangle$ [Eq.~(\ref{eq:H-BBQ-triangle})] 
can be defined on a triangular plaquette containing 
a triad of directors [cf.~Fig.~\ref{fig:AFNtriangle}]. 
These directors, which we will label {\sf A}, {\sf B} and {\sf C}, could in principle be located on the sites 
of the lattice, as is the case here, or on the bonds, as is the case in multiple spin exchange 
models relevant to thin films of $^3$He.


At the high symmetry ${\sf SU(3)}$ point, \mbox{$J_1=J_2=J$}, the Hamiltonian, 
$\mathcal{H}^{\sf BBQ}_\triangle$ [Eq.~(\ref{eq:averageH})], simplifies to,
\begin{align}
 \mathcal{H}_{\sf SU(3)}  
= J\left( 
|{\bf d}_{\sf A}\cdot\bar{{\bf d}}_{\sf B}|^2 
+ |{\bf d}_{\sf B}\cdot\bar{{\bf d}}_{\sf C}|^2 
+ |{\bf d}_{\sf C}\cdot\bar{{\bf d}}_{\sf A}|^2 
\right)  +3J.
\label{eq:H-SU3}
\end{align}
This can be minimised by requiring \mbox{${\bf d}_i.\bar{{\bf d}}_j=0$} on every bond, 
resulting in a 3-sublattice order in which neighbouring ${\bf d}$ 
vectors are orthogonal. 
There is no requirement that ${\bf d}$ should 
be real (or imaginary) and therefore the ground state manifold includes both quadrupolar, 
dipolar and mixed phases. 


One choice for the ground state of such a system is,
\begin{align}
{\bf d}_{\sf A}^{gs} = (1,0,0), \quad
{\bf d}_{\sf B}^{gs} = (0,1,0), \quad
{\bf d}_{\sf C}^{gs} = (0,0,1).
\label{eq:xyzgs}
\end{align}
This corresponds to an AFQ state in which the three directors lie along the principle axes, 
$({\sf x,y,z})$, and is illustrated in Fig.~\ref{fig:AFNtriangle}.


\begin{figure*}[ht]
\centering
\includegraphics[width=0.8\textwidth]{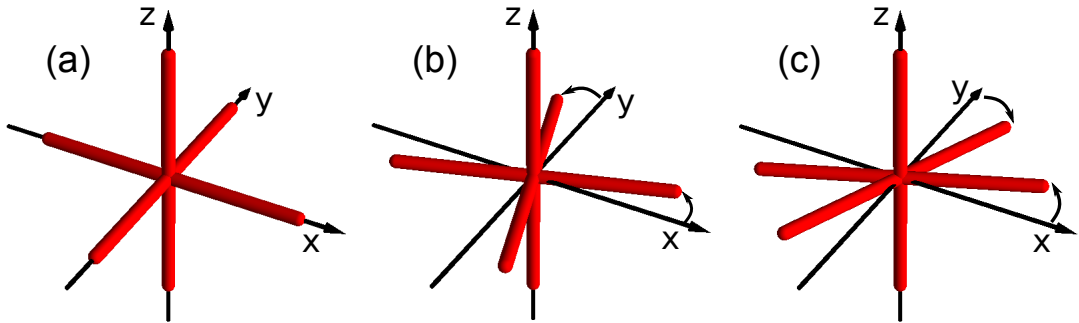}
\caption{\footnotesize{(Color online). 
Real  component of the complex director configurations for antiferroquadrupolar (AFQ) order 
on a triangular plaquette [see Fig.~\ref{fig:AFNtriangle}], showing the action of global rotations 
${\bf U}(\boldsymbol{\phi})$ and local rotations associated with the canting fields ${\bf l}$.
Directors (red cylinders) on the three sites of the plaquette are combined at the plaquette centre.
(a) shows the orthogonal ground state given in Eq.~(\ref{eq:xyzgs}).
(b) shows the result of acting on this particular ground state with ${\bf D}_\triangle(\phi_1,0,\dots)$ [see Eq.~(\ref{eq:Dmatrices})].
This performs a global rotation of the directors around the {\sf z}-axis, and a different orthogonal ground state is generated.
(c) shows the result of acting with ${\bf D}_\triangle(0,\dots,l_1^{\sf z},0,\dots )$, which is seen to rotate directors orientated along the $x$- and $y$-axes in opposite directions around the $z$-axis. 
In consequence the angle between the directors changes, and this costs energy according to the Hamiltonian $\langle \mathcal{H}^{\sf BBQ}_\triangle \rangle$ [Eq.~(\ref{eq:averageH})]. 
}}
\label{fig:AFQ-OPdynamics}
\end{figure*}


The Hamiltonian, Eq.~(\ref{eq:H-SU3}), is invariant under the global rotation 
\mbox{${\bf d} \rightarrow {\bf U} {\bf d}$}, provided that \mbox{${\bf U}^{-1}={\bf U}\dg$}, 
making clear the {\sf SU(3)} symmetry of the ground state.
However not all of the ground states generated by these rotations are physically distinct,
since one is free to fix the phase on a site.
There are in fact 6 distinct generators of rotations that transform the system between inequivalent ground states.
These are conveniently represented using six of the eight Gell-Mann matrices,
\begin{align}
\boldsymbol{\lambda}_1&=
\left(
\begin{array}{ccc}
0 & -i & 0  \\
i & 0  & 0 \\
0 & 0 &0
\end{array}
\right)
\qquad
\boldsymbol{\lambda}_2=
\left(
\begin{array}{ccc}
0 & 0 & -i  \\
0 & 0  & 0 \\
i & 0 &0
\end{array}
\right)
\nonumber \\
\boldsymbol{\lambda}_3 &=
\left(
\begin{array}{ccc}
0 & 0 & 0  \\
0 & 0  & -i \\
0 & i &0
\end{array}
\right)
\qquad
\boldsymbol{\lambda}_4=
\left(
\begin{array}{ccc}
0 & 1 & 0  \\
1 & 0  & 0 \\
0& 0 &0
\end{array}
\right)
\nonumber \\
\boldsymbol{\lambda}_5 &=
\left(
\begin{array}{ccc}
0 & 0 & 1  \\
0 & 0  & 0 \\
1 & 0 & 0
\end{array}
\right)
\qquad
\boldsymbol{\lambda}_6=
\left(
\begin{array}{ccc}
0 & 0 & 0  \\
0 & 0  & 1 \\
0 & 1 & 0
\end{array}
\right).
\end{align}
The other two Gell-man matrices are diagonal, and are not physically relevant, 
as they change the phase of the directors.


Starting from a particular ground state of the triangular plaquette, such as the one described in 
Eq.~(\ref{eq:xyzgs}), the global rotation matrix,
\begin{align}
{\bf U}(\boldsymbol{\phi})=\exp\left[{i\sum_{p=1}^6\lambda_p \phi_p}\right],
\label{eq:U}
\end{align}
can be used to explore all other possible ground state configurations,
where $\boldsymbol{\phi}=(\phi_1, \dots, \phi_6)$.
This matrix acts globally on all three ${\bf d}$-vectors, and thus preserves the angle between 
them in the complex vector space.
In consequence these rotations have a zero energy cost, and a real space illustration of this is 
shown in Fig.~\ref{fig:AFQ-OPdynamics}b.


The global rotations of the order parameter can be split into two categories.
In order to see this, it is useful to use the shorthand notation ${\bf U}_1={\bf U}(\phi_1,0,0,0,0,0)$ 
and similarly for ${\bf U}_2, \dots {\bf U}_6$.
The matrices ${\bf U}_1$, ${\bf U}_2$ and ${\bf U}_3$ perform rotations of the directors 
which are real in the sense that, if ${\bf d}$ is real [as in Eq.~(\ref{eq:xyzgs})], it will remain so 
under these transformations. 
Applied to the AFQ ground state, they act only to rotate the quadrupole moments.
However the matrices ${\bf U}_4$, ${\bf U}_5$ and ${\bf U}_6$ transform a real ${\bf d}$ vector into 
a complex one in such a way as to mix a dipolar component into the AFQ ground state.
We will return to this point below when classifying spin excitations.   


\subsection{Canting of a plaquette}
\label{sec:plaq-canting}


Our ultimate aim is to describe the long-wavelength, director-wave fluctuations about the 
`invisible' AFQ spin-nematic ground state --- the `waves in the unseen'.  
This involves canting of the director triad out of the orthogonal ground state.
%


A necessary first step, is to construct a matrix, ${\bf D}_\triangle$, that can be used to access any 
configuration of three ${\bf d}$-vectors on a triangular plaquette.
In order to do this, it is useful to introduce a second set of generators,
\begin{align}
\boldsymbol{\mu}_1&=
\left(
\begin{array}{ccc}
0 & -i & 0  \\
-i & 0  & 0 \\
0 & 0 &0
\end{array}
\right)
\qquad
\boldsymbol{\mu}_2=
\left(
\begin{array}{ccc}
0 & 0 & -i  \\
0 & 0  & 0 \\
-i & 0 &0
\end{array}
\right)
\nonumber \\
\boldsymbol{\mu}_3 &=
\left(
\begin{array}{ccc}
0 & 0 & 0  \\
0 & 0  & -i \\
0 & -i &0
\end{array}
\right)
\qquad
\boldsymbol{\mu}_4=
\left(
\begin{array}{ccc}
0 & 1 & 0  \\
-1 & 0  & 0 \\
0& 0 &0
\end{array}
\right)
\nonumber \\
\boldsymbol{\mu}_5 &=
\left(
\begin{array}{ccc}
0 & 0 & -1  \\
0 & 0  & 0 \\
1 & 0 & 0
\end{array}
\right)
\qquad
\boldsymbol{\mu}_6=
\left(
\begin{array}{ccc}
0 & 0 & 0  \\
0 & 0  & 1 \\
0 & -1 & 0
\end{array}
\right).
\end{align}
When these act on a triad of ${\bf d}$ vectors [see Fig.~\ref{fig:AFQ-OPdynamics}a], they change 
the angles between the vectors, thus changing the energy, according to Eq.~(\ref{eq:H-SU3}) 
[see Fig.~\ref{fig:AFQ-OPdynamics}c].
Any configuration of the three ${\bf d}$ vectors can be accessed from Eq.~(\ref{eq:xyzgs}) using,
\begin{align}
{\bf D}_\triangle(\boldsymbol{\phi},{\bf l})=\exp &\left[{i\sum_{p=1}^6\lambda_p \phi_p} +  
i\mu_1 l_1^{\sf z} + i\mu_2 l_1^{\sf y} +i\mu_3 l_1^{\sf x}  \right. \nonumber \\
&\qquad \left. + i\mu_4 l_2^{\sf z} +i \mu_5 l_2^{\sf y} +i\mu_6 l_2^{\sf x}
\right],
\label{eq:Dmatrices}
\end{align}
where the vector ${\bf l}$ is defined by, 
\begin{align}
{\bf l}=\left(
\begin{array}{c}
l^{\sf z} \\
l^{\sf x} \\
l^{\sf y} 
\end{array}
\right)
=\left(
\begin{array}{c}
l_1^{\sf z} + i l_2^{\sf z} \\
l_1^{\sf x} + i l_2^{\sf x} \\
l_1^{\sf y} + i l_2^{\sf y}
\end{array}
\right).
\label{eq:lvector}
\end{align}
This notation may appear unnatural at first sight, but will prove convenient for calculation.
A completely general configuration of the three ${\bf d}$ vectors is thus given by,
\begin{align}
{\bf d}_{\sf A}=
 {\bf D}_\triangle \cdot
\left(
\begin{array}{c}
1 \\
0  \\
0 
\end{array}
\right),
\
{\bf d}_{\sf B}  = {\bf D}_\triangle    \cdot
\left(
\begin{array}{c}
0 \\
1  \\
0 
\end{array}
\right),
\
{\bf d}_{\sf C} =
 {\bf D}_\triangle  \cdot
\left(
\begin{array}{c}
0 \\
0  \\
1 
\end{array}
\right)
\label{eq:direc}
\end{align}
%


We now make the assumption that the system has at least short-range order, and thus expand for small canting fields ${\bf l}$.
Retaining fields up to $\mathcal{O}({\bf l})$,
\begin{align}
{\bf d}_{\sf A} 
= {\bf U}  \cdot
\left(
\begin{array}{c}
1 \\
\bar{l}^{\sf z}  \\
l^{\sf y} 
\end{array}
\right),
\
{\bf d}_{\sf B}  
=
 {\bf U}  \cdot \left(
\begin{array}{c}
l^{\sf z}  \\
1 \\
\bar{l}^{\sf x} 
\end{array}
\right),
\
{\bf d}_{\sf C} 
= {\bf U}  \cdot
\left(
\begin{array}{c}
\bar{l}^{\sf y} \\
l^{\sf x} \\
1
\end{array}
\right),
\label{eq:approxdirec}
\end{align}
and it follows that the length and phase constraints of Eq.~(\ref{eq:lenthfix}) and Eq.~(\ref{eq:phasefix}) 
hold to $\mathcal{O}({\bf l}^2)$.


The eventual aim is to eliminate the canting fields ${\bf l}$ from the partition function by integration.
What will remain is a theory describing the dynamics of the order parameter matrix, ${\bf U}$, in terms of the variables 
$\boldsymbol{\phi}$.


\subsubsection{Continuum limit}

\begin{figure}[t]
\centering
\includegraphics[width=0.48\textwidth]{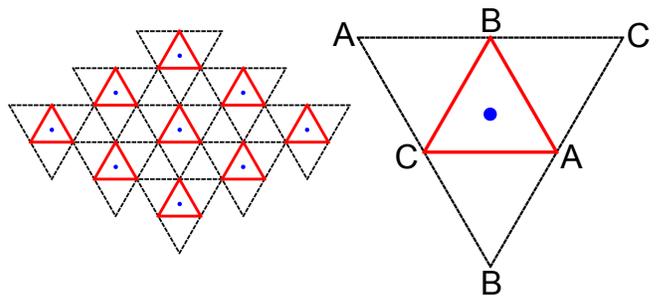}
\caption{\footnotesize{(Color online). 
The partitioning of the triangular lattice used in the derivation of the field theory for the 
3-sublattice antiferroquadrupolar (AFQ) spin-nematic state. 
The lattice is split into clusters containing 3 sites and 9 bonds, such that each cluster retains the 
point group symmetries of the lattice. 
Fields are defined at the centre of the clusters (blue dots), and the fact that the directors are associated 
with the vertices of the lattice is built into the continuum theory by a gradient expansion about this point.
}}
\label{fig:triangularlattice}
\end{figure}

We now consider how to pass from a lattice theory to a continuum theory of the AFQ state. 
The lattice can be partitioned into clusters based on 
triangular plaquettes, as shown in Fig.~\ref{fig:triangularlattice}). 
The director fields are defined at the centre of these clusters, and the physical 
location of the directors is taken into account by performing a gradient expansion.  
The continuum limit involves the assumption that physically 
interesting variation takes place on a lengthscale much larger than the lattice 
constant, $a$, and so gradients within the plaquette are small.


%
%


One of the requirements of a continuum field theory is that it should describe the dynamics of both the 
broken symmetry state {\it and} the nearby paramagnetic region, in which the order parameter is 
assumed to be locally robust but slowly varying over macroscopic length scales. 
It is therefore necessary to allow the fields to fluctuate in space and time,
\begin{align}
{\bf d}_{\sf A}({\bf r},\tau) &= {\bf U}({\bf r},\tau)
\left(
\begin{array}{c}
1 \\
\bar{l}^{\sf z}({\bf r},\tau)  \\
l^{\sf y}({\bf r},\tau) 
\end{array}
\right)
+\mathcal{O}({\bf l}^2),
\nonumber \\
{\bf d}_{\sf B}({\bf r},\tau)  &= {\bf U}({\bf r},\tau)
\left(
\begin{array}{c}
l^{\sf z}({\bf r},\tau)  \\
1 \\
\bar{l}^{\sf x}({\bf r},\tau) 
\end{array}
\right)
+\mathcal{O}({\bf l}^2),
\nonumber \\
{\bf d}_{\sf C}({\bf r},\tau) &= {\bf U} ({\bf r},\tau)
\left(
\begin{array}{c}
\bar{l}^{\sf y}({\bf r},\tau) \\
l^{\sf x}({\bf r},\tau) \\
1
\end{array}
\right)
+\mathcal{O}({\bf l}^2).
\label{eq:directors-as-U-times-l}
\end{align}


A useful parametrisation of the matrix $ {\bf U}$ is,
\begin{align}
 {\bf U}({\bf r},\tau)= \left(
\begin{array}{ccc}
n_{\sf A}^{\sf x}({\bf r},\tau) & n_{\sf B}^{\sf x}({\bf r},\tau) & n_{\sf C}^{\sf x}({\bf r},\tau)  \\
n_{\sf A}^{\sf y}({\bf r},\tau) & n_{\sf B}^{\sf y}({\bf r},\tau)  & n_{\sf C}^{\sf y}({\bf r},\tau) \\
n_{\sf A}^{\sf z}({\bf r},\tau) & n_{\sf B}^{\sf z}({\bf r},\tau) & n_{\sf C}^{\sf z}({\bf r},\tau)
\end{array}
\label{eq:Uvsn}
\right),
\end{align}
where the complex fields ${\bf n}_{\sf i}({\bf r},\tau)$, with ${\sf i}=\{ {\sf A}, {\sf B}, {\sf C}  \}$, inherit the length and phase constraints 
of the ${\bf d}$ vectors (Eq.~(\ref{eq:lenthfix}) and Eq.~(\ref{eq:phasefix})),
\begin{align}
{\bf n}_{\sf i}\cdot\bar{{\bf n}}_{\sf i} =1, \quad
{\bf n}_{\sf i}^2-\bar{{\bf n}}_{\sf i}^2=0,
\label{eq:nconstraints1}
\end{align}
and are also required to be orthogonal to one another according to,
\begin{align}
&{\bf n}_{\sf i}\cdot\bar{{\bf n}}_{\sf j}=0, \quad
{\sf i} \neq {\sf j} .
\label{eq:nconstraints2}
\end{align}
The apparent 18 degrees of freedom of the ${\bf n}_{\sf i}$ fields is reduced  to 6 by the 12 
constraints, as expected. 
The reason that the parametrisation in terms of 
${\bf n}_{\sf i}({\bf r},\tau)$,
is useful is that these fields are mutually orthogonal, of unit length, and of fixed phase, 
and can therefore be interpreted as a ground-state director configuration. 
In consequence there are two equivalent formulations of the field theory: in terms of the rotation matrix 
$ {\bf U}({\bf r},\tau)$; or in terms of the fields ${\bf n}_{\sf i}({\bf r},
\tau)$.
We will make use of both in what follows.


Differentiating the constraints, Eq.~(\ref{eq:nconstraints1}) and Eq.~(\ref{eq:nconstraints2}), leads to the relations,
\begin{align}
&{\bf n}_{\sf i}\cdot\partial_\lambda \bar{{\bf n}}_{\sf i}=-\bar{{\bf n}}_{\sf i}. \partial_ \lambda {\bf n}_{\sf i}, \quad 
{\bf n}_{\sf i}\cdot\partial_ \lambda {\bf n}_{\sf i}=\bar{{\bf n}}_{\sf i}. \partial_ \lambda \bar{{\bf n}}_{\sf i}, \nonumber \\
& \qquad {\bf n}_{\sf i}\cdot\partial_ \lambda \bar{{\bf n}}_{\sf j}=-\bar{{\bf n}}_{\sf j}. \partial_ \lambda {\bf n}_{\sf i}, \quad
{\sf i} \neq {\sf j} ,
\label{eq:nconditions}
\end{align}
where the partial derivative $\partial_\lambda$ can be with respect to any space-time variable.
These relations prove very useful for simplifying subsequent expressions.   


The partition function can be written in terms of a functional integral over all director configurations,
\begin{align}
\mathcal{Z}_\triangle^{\sf SU(3)} = \int \mathcal{D}[{\bf d}] e^{-\mathcal{S}^{\sf SU(3)}_\triangle[{\bf d}]},
\end{align}
where $\mathcal{S}^{\sf SU(3)}_\triangle[{\bf d}]$ is the Euclidean action and the integration 
measure $\mathcal{D}[{\bf d}]$ includes the delta function constraints on the length 
and phase of the director. 
The action can be split into Hamiltonian and kinetic terms,
\begin{align}
\mathcal{S}^{\sf SU(3)}_\triangle = \mathcal{S}_{\sf kin}+\mathcal{S}_{\mathcal{H}[{\sf SU(3)}]},
\end{align}
where $\mathcal{S}_{\sf kin}$ is a dynamic, geometric-phase term and 
$\mathcal{S}_{\mathcal{H}[{\sf SU(3)}]}$ accounts for the energy cost of static 
director configurations at the ${\sf SU(3)}$ point.


\subsubsection{The Hamiltonian term}


The energy cost of a particular static configuration of directors is given by Eq.~(\ref{eq:averageH}). 
In principle, the Hamiltonian term in the action, 
$ \mathcal{S}_\mathcal{H}$, takes into account all static configurations of directors. 
However, we make the approximation that only those with a slow 
spatial variation are important.


The Hamiltonian term is given by,
\begin{align}
\mathcal{S}_{\mathcal{H}[\sf{SU(3)}]} 
    &= \int_0^\beta d\tau   \mathcal{H}_{\sf SU(3)}   \nonumber\\
    &= \frac{2}{3\sqrt{3}a^2} 
    \int_0^\beta d\tau \int d^2r 
    \mathcal{H}_{\sf SU(3),clus},
\label{eq:S-kin}
\end{align}
where $ \mathcal{H}_{\sf SU(3), clus}$ refers to the Hamiltonian for a single cluster, 
and the numerical prefactor is related to the area of the cluster.


The gradient expansion of the fields in terms of the small parameter $a$ is given by,
 \begin{align}
{\bf d}_j({\bf r} + \boldsymbol{\epsilon}_i, \tau) & = 
{\bf d}_j({\bf r}, \tau) 
+ a  ( \boldsymbol{\epsilon}_i \cdot \nabla) {\bf d}_j({\bf r}, \tau) \nonumber \\ &
+ \frac{a^2}{2!}( \boldsymbol{\epsilon}_i \cdot \nabla)^2 {\bf d}_j({\bf r}, \tau) 
+ \mathcal{O}(a^3),
\label{eq:gradexpansion}
 \end{align}
 where $\boldsymbol{\epsilon}_i$ is the vector connecting the centre of the cluster to the 
 lattice sites within it  (cf. Fig.~\ref{fig:triangularlattice}). 
  

Expanding the Hamiltonian to second order in the lattice parameter $a$ gives,
 \begin{align}
& \mathcal{H}_{\sf SU(3), clus}  \approx 
  3J\left(  
  |\bar{\bf d}_{\sf A}({\bf r},\tau)\cdot{\bf d}_{\sf B}({\bf r},\tau)|^2 \right. \nonumber \\
&\left. \qquad +|\bar{\bf d}_{\sf B}({\bf r},\tau)\cdot{\bf d}_{\sf C}({\bf r},\tau)|^2 +|\bar{\bf d}_{\sf C}({\bf r},\tau)\cdot{\bf d}_{\sf A}({\bf r},\tau)|^2 \right) 
\nonumber \\
 &+\frac{3Ja^2}{2} \sum_{\lambda ={\sf x,y}} \left[ |\bar{{\bf d}}_{\sf A}\cdot\partial_ \lambda {\bf d}_{\sf B}|^2 
  + |\bar{{\bf d}}_{\sf B}\cdot\partial_\lambda {\bf d}_{\sf C}|^2
 +|\bar{{\bf d}}_{\sf C}\cdot\partial_\lambda {\bf d}_{\sf A}|^2  \right]. 
\end{align}
The first term in this expression vanishes if the system is an AFQ ground state.
Fluctuations about this can be expanded in terms of the canting field ${\bf l}$ using,
 \begin{align}
\bar{{\bf d}}_{\sf A}({\bf r},\tau) \cdot {\bf d}_{\sf B}({\bf r},\tau) &\approx 2l^{\sf z}({\bf r},\tau) \nonumber \\
\bar{{\bf d}}_{\sf B}({\bf r},\tau) \cdot {\bf d}_{\sf C}({\bf r},\tau) & \approx 2l^{\sf x}({\bf r},\tau) \nonumber \\
\bar{{\bf d}}_{\sf C}({\bf r},\tau) \cdot {\bf d}_{\sf A}({\bf r},\tau) & \approx 2l^{\sf y}({\bf r},\tau).
\end{align}
Since the gradient terms are already $\mathcal{O}(a^2)$, the fields ${\bf d}({\bf r},\tau)$ can be replaced by the orthogonal fields ${\bf n}({\bf r},\tau)
$, giving the Hamiltonian,
 \begin{align}
& \mathcal{H}_{\sf SU(3), clus} \approx 
  12J  {\bf l} \cdot \bar{{\bf l}}
   \nonumber \\
 &+\frac{3Ja^2}{2} \sum_{\lambda ={\sf x,y}} \left[ |\bar{{\bf n}}_{\sf A}\cdot\partial_ \lambda {\bf n}_{\sf B}|^2
 +|\bar{{\bf n}}_{\sf B}\cdot\partial_ \lambda {\bf n}_{\sf C}|^2
 +|\bar{{\bf n}}_{\sf C}\cdot\partial_ \lambda {\bf n}_{\sf A}|^2  \right].
\end{align}


\subsubsection{The kinetic term}


The action describing long wave-length fluctuations of the AFQ state 
also contains a kinetic energy term. 
This is quantum-mechanical in origin, and a consequence of the overcompleteness 
of the coherent states used to represent spin configurations.
At a semiclassical level it describes the rotational motion of the directors, and 
can therefore be interpreted as a geometrical phase.
For a more detailed explanation we refer the interested reader to the chapters on spin path integrals in 
[\onlinecite{fradkin,auerbach}].


The contribution of the kinetic term to the action is,
 \begin{align}
\mathcal{S}_{\sf kin} \approx  \int_0^\beta d\tau \frac{2}{3\sqrt{3}a^2} \int d^2r 
\sum_{\sf i}
 \bar{{\bf d}}_{\sf i}\cdot\partial_\tau {\bf d}_{\sf i} 
 \end{align}
where spatial gradient terms have been ignored. 
To first order in the canting field ${\bf l}$,
 \begin{align}
\sum_{\sf i}
 \bar{{\bf d}}_{\sf i}\cdot\partial_\tau {\bf d}_{\sf i} 
\approx 
 \mathrm{Tr}[{\bf U}\dg\cdot\partial_\tau {\bf U}] + 2\left[ {\bf s} \cdot {\bf l} -  \bar{{\bf s}} \cdot \bar{{\bf l}} \right],
\end{align}
where the complex field ${\bf s}({\bf r},\tau)$ is defined as,
\begin{align}
 {\bf s}=
  \left(
\begin{array}{c}
 ({\bf U}\dg\cdot\partial_\tau {\bf U})_{21} \\
 ({\bf U}\dg\cdot\partial_\tau {\bf U})_{32} \\
 ({\bf U}\dg\cdot\partial_\tau {\bf U})_{13}
\end{array}
\right)
= \left(
\begin{array}{c}
\bar{{\bf n}}_{\sf B}\cdot\partial_\tau {\bf n}_{\sf A} \\
\bar{{\bf n}}_{\sf C}\cdot\partial_ \tau {\bf n}_{\sf B} \\
\bar{{\bf n}}_{\sf A}\cdot\partial_ \tau {\bf n}_{\sf C} 
\end{array}
\right).
\label{eq:sfield}
\end{align}
The kinetic term gives an imaginary contribution to the Euclidean Lagrangian. 
Derivatives of the field ${\bf l}$ vanish, since they are total derivatives 
and can therefore be converted to a vanishing surface integral.


\subsubsection{Integrating out fluctuations}


Having derived an action for long-wavelength fluctuations of the AFQ state, 
the task which remains is to eliminate the canting fields ${\bf l}({\bf r},\tau)$,  
so as to arrive at an action written entirely in terms of the order parameter 
${\bf n}({\bf r},\tau)$.
Taking into account both potential and kinetic energy terms
in the Hamiltonian, we start from the partition function,

\begin{align}
\mathcal{Z}_\triangle^{\sf SU(3)} 
   \propto & \int \prod_{{\sf i}\neq j}
            \mathcal{D}{\bf n}_{\sf i} 
            \mathcal{D}\bar{{\bf n}}_{\sf i} 
            \mathcal{D}{\bf l} 
            \mathcal{D}\bar{{\bf l}} 
\ \delta({\bf n}_{\sf i}\cdot\bar{{\bf n}}_{\sf i}-1) 
\ \delta({\bf n}_{\sf i}^2-\bar{{\bf n}}_{\sf i}^2) \nonumber \\
&\delta({\bf n}_{\sf i}\cdot\bar{{\bf n}}_{\sf j}) 
\ e^{-\mathcal{S}^{\sf SU(3)}_\triangle[{\bf n}_{\sf A},\bar{{\bf n}}_{\sf A},{\bf n}_{\sf B},\bar{{\bf n}}_{\sf B},{\bf n}_{\sf C},\bar{{\bf n}}_{\sf C},{\bf l},
\bar{{\bf l}}]},
\end{align}
where the action,
\begin{align}
&\mathcal{S}^{\sf SU(3)}_\triangle[{\bf n}_{\sf A},\bar{{\bf n}}_{\sf A},{\bf n}_{\sf B},\bar{{\bf n}}_{\sf B},{\bf n}_{\sf C},\bar{{\bf n}}_{\sf C},{\bf l},\bar{{\bf l}}] 
\nonumber \\
&\qquad \qquad=  \int_0^\beta d\tau \frac{2}{3\sqrt{3}a^2} \int d^2r \mathcal{L}_\triangle^{\sf SU(3)},
\end{align}
is written in terms of the Lagrangian,
\begin{align}
& \mathcal{L}_\triangle^{\sf SU(3)} \approx 
   \mathrm{Tr}[{\bf U}\dg\cdot\partial_\tau {\bf U}] +  2\left[ {\bf s} \cdot {\bf l} -  \bar{{\bf s}} \cdot \bar{{\bf l}} \right]
  + 12J  {\bf l} \cdot \bar{{\bf l}}
   \nonumber \\
 &+\frac{3Ja^2}{2} \sum_{\lambda ={\sf x,y}} \left[ |\bar{{\bf n}}_{\sf A}\cdot\partial_ \lambda {\bf n}_{\sf B}|^2
 +|\bar{{\bf n}}_{\sf B}\cdot\partial_ \lambda {\bf n}_{\sf C}|^2
 +|\bar{{\bf n}}_{\sf C}\cdot\partial_ \lambda {\bf n}_{\sf A}|^2  \right].
\label{eq:Ll1l2}
\end{align}


The canting fields ${\bf l}$ and $\bar{{\bf l}}$ enter the Lagrangian at a quadratic level and can therefore 
be eliminated via a Gaussian integral, or,  equivalently, using the steepest-descent approximation. 
This process is slightly simpler if the two fields are decoupled by the linear transformation,
\begin{align}
{\bf l} = {\bf l}_1 +i {\bf l}_2, \qquad \bar{{\bf l}} = {\bf l}_1 -i {\bf l}_2,
\label{eq:fielddiag}
\end{align}
where ${\bf l}_1$ and ${\bf l}_2$ are real.
Taking functional derivatives with respect to these fields gives,
\begin{align}
\frac{\delta \mathcal{L}_\triangle^{\sf SU(3)}}{\delta {\bf l}_1} &\approx 2({\bf s}-\bar{{\bf s}}) +24J {\bf l}_1 \approx  0 \nonumber \\
\frac{\delta \mathcal{L}_\triangle^{\sf SU(3)}}{\delta {\bf l}_2} & \approx 2i({\bf s}+\bar{{\bf s}}) +24J {\bf l}_2 \approx  0,
\end{align}
and  these equations are resolved as,
\begin{align}
{\bf l}_1 &\approx -\frac{1}{12J} ({\bf s}-\bar{{\bf s}}) \nonumber \\
{\bf l}_2 &\approx -\frac{i}{12J} ({\bf s}+\bar{{\bf s}}).
\label{eq:lintermsofs}
\end{align}


At this point it is helpful to introduce a `director stiffness',
\begin{align}
\qquad \rho_{\sf d}=Ja^2,
\end{align}
describing the energy cost of twisting the order parameter, 
and the generalised susceptibility,
\begin{align}
\chi_\perp = \frac{2}{9J},
\end{align}
associated with fluctuations of the canting field ${\bf l}$.   


Substituting the canting fields, Eq.~(\ref{eq:lintermsofs}), into the Lagrangian, Eq.~(\ref{eq:Ll1l2}), and 
using Eq.~(\ref{eq:sfield}) and Eq.~(\ref{eq:Uvsn}) to re-express this in terms of the fields 
${\bf n}_{\sf i}$, we arrive at
\begin{align}
& \mathcal{S}^{\sf SU(3)}_\triangle[
{\bf n}_{\sf A},
\bar{{\bf n}}_{\sf A},
{\bf n}_{\sf B},
\bar{{\bf n}}_{\sf B},
{\bf n}_{\sf C},
\bar{{\bf n}}_{\sf C}] 
\nonumber  \\
& =  \frac{1}{\sqrt{3}a^2} 
\int_0^\beta d\tau \int d^2r \left\{ 
  \frac{2}{3}
\sum_{\sf i} \bar{{\bf n}}_{\sf i}\cdot\partial_\tau {\bf n}_{\sf i}  \right.
\nonumber  \\
& +  \chi_\perp \left[ |\bar{{\bf n}}_{\sf A}\cdot\partial_ \tau {\bf n}_{\sf B}|^2
 +|\bar{{\bf n}}_{\sf B}\cdot\partial_ \tau {\bf n}_{\sf C}|^2
 +|\bar{{\bf n}}_{\sf C}\cdot\partial_ \tau {\bf n}_{\sf A}|^2  \right]  
 \nonumber \\
& \left. + \rho_{\sf d} \sum_{\lambda ={\sf x,y}} 
 \left[ |\bar{{\bf n}}_{\sf A}\cdot\partial_ \lambda {\bf n}_{\sf B}|^2
 +|\bar{{\bf n}}_{\sf B}\cdot\partial_ \lambda {\bf n}_{\sf C}|^2
 +|\bar{{\bf n}}_{\sf C}\cdot\partial_ \lambda {\bf n}_{\sf A}|^2  \right]
  \right\},
 \nonumber \\
\label{eq:SSU3}
\end{align}
with associated partition function,
\begin{align}
\mathcal{Z}_\triangle^{\sf SU(3)} 
   \propto & \int \prod_{{\sf i}\neq j}
            \mathcal{D}{\bf n}_{\sf i} 
            \mathcal{D}\bar{{\bf n}}_{\sf i} 
\ \delta({\bf n}_{\sf i}\cdot\bar{{\bf n}}_{\sf i}-1) 
\ \delta({\bf n}_{\sf i}^2-\bar{{\bf n}}_{\sf i}^2) \nonumber \\
&\delta({\bf n}_{\sf i}\cdot\bar{{\bf n}}_{\sf j}) 
\ e^{-\mathcal{S}^{\sf SU(3)}_\triangle[{\bf n}_{\sf A},\bar{{\bf n}}_{\sf A},{\bf n}_{\sf B},\bar{{\bf n}}_{\sf B},{\bf n}_{\sf C},\bar{{\bf n}}_{\sf C}]},
\label{eq:ZSU3}
\end{align}
where the canting fields have been eliminated at a Gaussian level.


Equivalently, Eq.~(\ref{eq:Uvsn}) can be used to write the action, Eq.~(\ref{eq:SSU3}), in terms of the 
unitary matrices, ${\bf U}({\bf r},\tau)$, as,
\begin{align}
&\mathcal{S}^{\sf SU(3)}_\triangle[{\bf U}] =   \frac{1}{2\sqrt{3}a^2} \int_0^\beta d\tau \int d^2r
\left\{ \frac{4}{3}\mathrm{Tr}[{\bf U}\dg\cdot\partial_\tau {\bf U}] \right.\nonumber \\
&\qquad + \chi_\perp \left[  \mathrm{Tr}[\partial_\tau{\bf U}\dg\cdot\partial_\tau {\bf U}] 
-\sum_m \left| [{\bf U}\dg\cdot\partial_\tau {\bf U}]_{mm} \right|^2 \right] 
\nonumber \\
& \qquad \left. + \rho_{\sf d} \sum_{\lambda={\sf x,y}}  
\left[  
\mathrm{Tr}[\partial_ \lambda{\bf U}\dg\cdot\partial_ \lambda {\bf U}] 
-\sum_m \left| [{\bf U}\dg\cdot\partial_ \lambda {\bf U}]_{mm} \right|^2 
\right]
\right\}
\label{eq:SSU3Umat}
\end{align}
where $m=\{1,2,3 \}$ labels matrix elements. 
This formulation of the action is further removed from the physical state than Eq.~(\ref{eq:SSU3}), 
but makes explicit the ${\sf SU}(3)$ symmetry of the Hamiltonian.


\subsubsection{Linearising the order parameter fields}
\label{sec:linearising-SU3}


The physical nature of the excitations of the AFQ state --- and in particular the division into 
quadrupole-wave and spin-wave modes --- is easier to understand once the action describing 
them has been linearized.
This can be achieved by expanding fluctuations about the AFQ ground state 
to leading order in $\boldsymbol{\phi}$.
We will consider in detail the interaction of the $\boldsymbol{\phi}$ fields in a future publication\cite{smerald-unpub}.

After linearization, the unitary matrix field ${\bf U}({\bf r}, \tau)$ [Eq.~(\ref{eq:U})] is approximated by,
\begin{align}
 {\bf U}({\bf r},\tau)  \approx \left(
\begin{array}{ccc}
1 & \phi_1+i\phi_4 & -\phi_2+i\phi_5  \\
 -\phi_1+i\phi_4 & 1  & \phi_3+i\phi_6 \\
\phi_2+i\phi_5 & -\phi_3+i\phi_6 & 1
\end{array}
\right),
\label{eq:linearized-U}
\end{align}
where the angular variables \mbox{$\phi_p=\phi_p({\bf r},\tau)$} fluctuate in both 
space and time.
It follows from Eq.~(\ref{eq:Uvsn}) that the fields ${\bf n}_{\sf i}({\bf r},\tau)$ 
are given by,
\begin{align}
{\bf n}_{\sf A}({\bf r},\tau)&\approx
\left(
\begin{array}{c}
1 \\
 -\phi_1+i\phi_4 \\
\phi_2+i\phi_5
\end{array}
\right),
\nonumber \\
{\bf n}_{\sf B}({\bf r},\tau)&\approx
\left(
\begin{array}{c}
\phi_1+i\phi_4 \\
1 \\
-\phi_3+i\phi_6
\end{array}
\right),
\nonumber \\
{\bf n}_{\sf C}({\bf r},\tau)&\approx
\left(
\begin{array}{c}
-\phi_2+i\phi_5 \\
\phi_3+i\phi_6 \\
1
\end{array}
\right).
\label{eq:linearisation}
\end{align}
making explicit that the fields \mbox{$\phi_p({\bf r},\tau)$} have a simple interpretation in terms 
of small, local angles of rotation away from the direction of spontaneous symmetry breaking.


Eq.~(\ref{eq:SQduvmatrix}) can now be used to reconstruct the fluctuating dipolar and quadrupolar 
moments on each sublattice.  
To leading order in \mbox{$\phi_p({\bf r},\tau)$}, these can be written as,
\begin{align}
{\bf S}_{\sf A}\approx
2\left(
\begin{array}{c}
0 \\
-\phi_5 \\
\phi_4
\end{array}
\right),
\
{\bf S}_{\sf B}\approx
2\left(
\begin{array}{c}
\phi_6 \\
0 \\
-\phi_4
\end{array}
\right),
\
{\bf S}_{\sf C}\approx
2\left(
\begin{array}{c}
-\phi_6 \\
\phi_5 \\
0
\end{array}
\right),
\end{align}
and
\begin{align}
{\bf Q}_{\sf A}\approx
\left(
\begin{array}{c}
-1 \\
1/\sqrt{3} \\
-2\phi_1 \\
0 \\
-2\phi_2
\end{array}
\right),
\
{\bf Q}_{\sf B}\approx
\left(
\begin{array}{c}
1 \\
1/\sqrt{3} \\
2\phi_1 \\
-2\phi_3 \\
0
\end{array}
\right),
\
{\bf Q}_{\sf C}\approx
\left(
\begin{array}{c}
0 \\
-2/\sqrt{3} \\
0 \\
2\phi_3 \\
2\phi_2
\end{array}
\right).
\label{eq:SQmoments}
\end{align}
This shows that the fields $\phi_1$, $\phi_2$ and $\phi_3$ are primarily associated with fluctuations 
of the quadrupole moments, and so justifies the name ``quadrupole-waves''. 
Since the fields $\phi_4$, $\phi_5$ and $\phi_6$ are primarily associated with {\it transverse} 
fluctuations of the dipole moments, we refer to them as ``spin-waves''.   
In Section~\ref{sec:SU2} we extend this analysis to also include  
time derivatives of the $\phi$ fields.   
The supplemental material contains animations showing the nature of the quadrupole-wave\cite{AFQ-Qwave-animation} and spin-wave\cite{AFQ-Qwave-animation} modes that follow from Eq.~(\ref{eq:SQmoments}). 


Linearizing the action $\mathcal{S}^{\sf SU(3)}_\triangle[{\bf U}]$ [Eq.~(\ref{eq:SSU3Umat})]
also enables us to eliminate the delta function constraints
from the partition function $\mathcal{Z}_\triangle^{\sf SU(3)}$ [Eq.~(\ref{eq:ZSU3})], to give
\begin{align}
\mathcal{Z}_\triangle^{\sf SU(3)} \propto &\int \mathcal{D}\boldsymbol{\phi}
e^{-\mathcal{S}^{\sf SU(3)}_\triangle[\boldsymbol{\phi}]},
\end{align}
where the linearised action is,
\begin{align}
&\mathcal{S}^{\sf SU(3)}_\triangle[\boldsymbol{\phi}] \approx \frac{1}{\sqrt{3}a^2} \int_0^\beta d\tau  \int d^2r
\sum_{p=1}^6 \nonumber  \\
& \qquad \left[ 
\chi_\perp (\partial_\tau \phi_p)^2 +\rho_{\sf d} \sum_{\lambda={\sf x,y}} (\partial_\lambda \phi_p)^2
\right].
\label{eq:linSU3action}
\end{align}

\begin{figure*}[ht]
\centering
\includegraphics[width=0.8\textwidth]{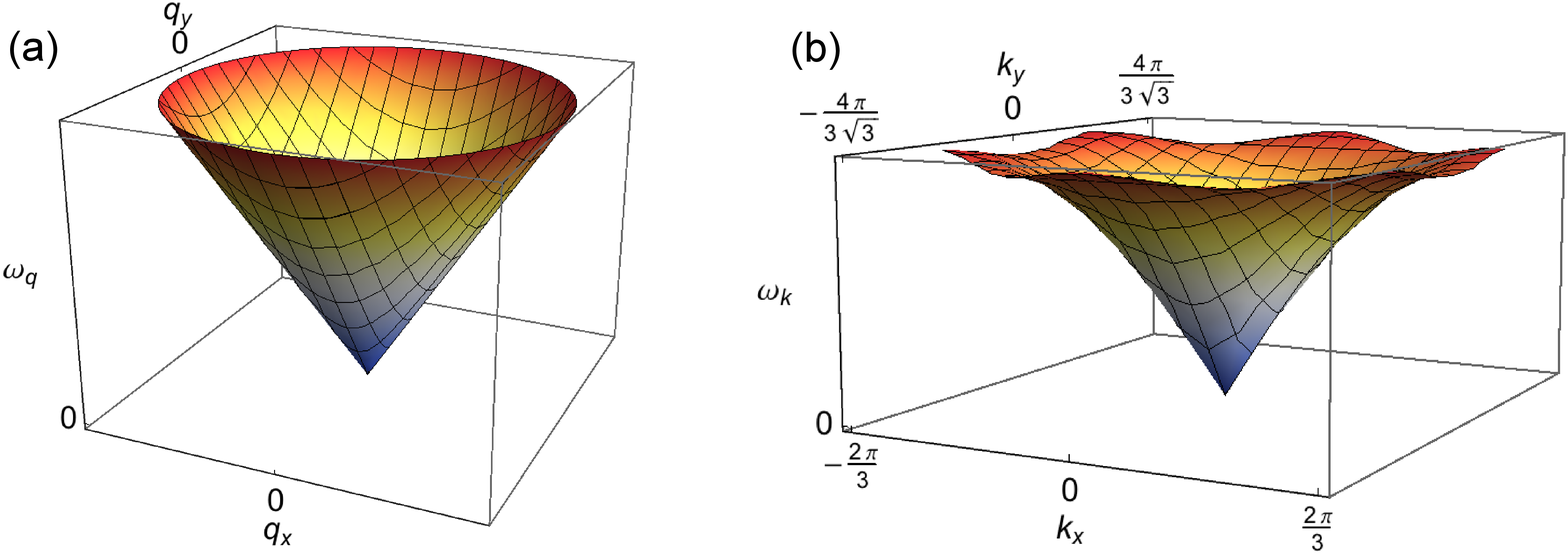}
\caption{\footnotesize{(Color online). 
The dispersion of magnetic excitations at the ${\sf SU(3)}$ point.
(a) prediction of the continuum field theory $\mathcal{S}^{\sf SU(3)}_\triangle$~[Eq.~(\ref{eq:linSU3action})].
(b) prediction of the \mbox{spin-1} 
bilinear-biquadratic (BBQ) model on a triangular lattice, ${\mathcal H}^{\sf BBQ}_\triangle$~[Eq.~(\ref{eq:H-BBQ-triangle})] for $J_1$=$J_2$. 
Approaching the ordering vector ${\bf k}={\bf k}_{\sf M}$ (${\bf q}=0$), the continuum theory and the lattice theory match exactly.
At this high-symmetry point there is a 6-fold degenerate branch of linearly-dispersive, 
gapless excitations.
These can be split into 3 modes that primarily describe fluctuations of quadrupole moments 
(quadrupole waves) and 3 that primarily describe fluctuations of dynamically generated 
dipole moments (spin waves).
These 3 spin-wave fields become gapped on entering the antiferroquadrupolar (AFQ) 
phase bordering the {\sf SU(3)} point.
}}
\label{fig:dispersionSU3}
\end{figure*}
\begin{figure}[h]
\centering
\includegraphics[width=0.35\textwidth]{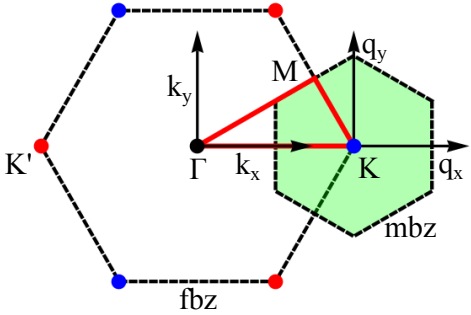}
\caption{\footnotesize{(Color online). 
The full Brillouin zone (fbz) of the triangular lattice, together with 
the reduced magnetic Brillouin zone (mbz) for 3-sublattice order. 
Important symmetry points are labelled 
$\Gamma$ [${\bf k}_\Gamma=(0,0)$], 
${\sf M}$ [${\bf k}_{\sf M}=(2\pi/3,\pi/\sqrt{3})$], 
${\sf K}$ [\mbox{${\bf k}_{\sf K}=(4\pi/3,0)$}] and 
${\sf K}^\prime$ [${\bf k}_{-{\sf K}}=-{\bf k}_{\sf K}$].
In the field theory for the 3-sublattice  antiferroquadrupolar (AFQ) 
state, the $\Gamma$ and ${\sf K}^\prime$ points are folded onto the 
${\sf K}$ point, and the wavevector ${\bf q}$ measures the deviation 
from this point. 
The circuit in reciprocal space $\Gamma$-${\sf K}$-${\sf M}$-$\Gamma$ 
followed when plotting the inelastic neutron 
scattering intensity in Fig.~\ref{fig:3SLAFQ-chiS}
is indicated in red.
}}
\label{fig:mbz}
\end{figure}

At this level of approximation the equations of motion for each field are independent of one another 
and given by,
\begin{align}
\left[ \chi_\perp \partial_\tau^2 +\rho_{\sf d} \partial_{\sf x}^2 +\rho_{\sf d} \partial_{\sf y}^2 \right] \phi_p=0.
\end{align}
These can be solved by the ansatz,
\begin{align}
\phi_p = A_p e^{i{\bf q}.{\bf r}+\omega_{\bf q}\tau},
\end{align}
and in consequence the dispersion (shown in Fig.~\ref{fig:dispersionSU3}) is,
\begin{align}
\omega_{\bf q} = \sqrt{\frac{\rho_{\sf d}}{\chi_\perp}} |{\bf q}| = v|{\bf q}|,
\end{align}
with the director-wave velocity,
\begin{align}
 v = \sqrt{\frac{\rho_{\sf d}}{\chi_\perp}} = \frac{3Ja}{\sqrt{2}}.
 \label{eq:SU3veloc}
\end{align}
Here the vector ${\bf q}$ measures the distance in reciprocal space from the centre of the magnetic 
Brillouin zone (mbz), which is centred on the ${\sf K}$ point, 
\mbox{${\bf k}_{\sf K} = (4\pi/3, 0)$}, 
as shown in Fig.~\ref{fig:mbz}.


Thus, at the ${\sf SU(3)}$ point,  there are 6 gapless excitations, which 
disperse linearly with the same velocity, regardless of whether they have spin-wave or 
quadrupole-wave character.   
This reflects the large ground state manifold at the ${\sf SU(3)}$ point, which consists of all 3-sublattice 
orthogonal arrangements of the ${\bf d}$ vectors, and therefore includes both the AFQ and AFM 
states [cf. Fig.~\ref{fig:phasediag}].  
In Section~\ref{sec:SU2} we show that, as $J_2$ is increased and dipolar order becomes energetically 
unfavourable, only three linearly dispersing modes remain --- the quadrupole-wave modes, which 
are the Goldstone modes of AFQ order.


We note that Tsunetsugu and Arikawa\cite{tsunetsugu06,tsunetsugu07} have previously determined 
the dispersion of Eq.~(\ref{eq:H-BBQ-triangle}) in the AFQ phase using a linearised 
``flavour-wave'' theory.
At the high symmetry ${\sf SU(3)}$ point they find,
\begin{align}
\omega_{\bf k} =3J\sqrt{1- |\gamma_{\bf k}|^2 },
\label{eq:dwaveSU3disp}
\end{align}
where,
\begin{align}
\gamma_{\bf k} 
= \frac{1}{3}\left( e^{ik_{\sf x}a} +2e^{\frac{-ik_{\sf x}a}{2}} \cos\frac{\sqrt{3}k_{\sf y}a}{2} \right).
\label{eq:gamma}
\end{align} 
As ${\bf k}\rightarrow 0$ the limiting value of Eq.~(\ref{eq:dwaveSU3disp}) is $\omega_{\bf k}\approx v|{\bf k}|$, 
where the velocity $v = 3Ja/\sqrt{2}$ is identical to the one predicted by the field theory [Eq.~(\ref{eq:SU3veloc})].


\subsection{Continuum theory away from the {\sf SU(3)} point}
\label{sec:SU2}


\subsubsection{Symmetry breaking terms}


The ${\sf SU(3)}$ point of ${\mathcal H}_\triangle^{\sf BBQ}$ [Eq.~(\ref{eq:H-BBQ-triangle})],  $J_2 = J_1 = J$, has 
an artificially high symmetry.
For $J_2 > J_1$ the symmetry of ${\mathcal H}_\triangle^{\sf BBQ}$ is reduced to ${\sf SU(2)}$, with
important implications for the excitations of the AFQ state.
In what follows, we construct a continuum field theory for the AFQ phase by perturbing 
away from the ${\sf SU(3)}$ point.
The most significant change, required for the stability of AFQ order, is the opening of a gap to 
the 3 spin-wave modes.  


Following the notation of Section~\ref{sec:SU3}, 
the Hamiltonian ${\mathcal H}^{\sf BBQ}_\triangle$ [Eq.~(\ref{eq:H-BBQ-triangle})] can be written 
\begin{align}
 {\mathcal H}_{\sf BBQ}  
   =  \mathcal{H}_{\sf SU(3)}  +  \Delta\mathcal{H}_{\sf SU(2)} 
\end{align}
where 
\begin{align}
\Delta\mathcal{H}_{\sf SU(2)} = (J_2-J_1)\sum_{\langle ij\rangle}|{\bf d}_i\cdot{\bf d}_j|^2,
\end{align}
and $\mathcal{H}_{\sf SU(3)} $ is defined by Eq.~(\ref{eq:H-SU3}).
In order to develop a pertubative expansion around the high-symmetry ${\sf SU(3)}$ point, 
we make the assumption that $J_2 - J_1 \ll J_1,J_2$. 
This assumption breaks down  for $\theta \to \pi/2$, and places a limit on the range of 
wavelengths for which the sigma-model description developed in this Section is a valid
description of ${\mathcal H}^{\sf BBQ}_\triangle$~[Eq.~(\ref{eq:H-BBQ-triangle})].  


The kinetic term in the action $\mathcal{S}_{\sf kin}$ [Eq.~(\ref{eq:S-kin})] is unchanged since it is 
a property of the coherent state representation of the spin states, not of the Hamiltonian. 
The change to the Hamiltonian term in the action $\mathcal{S}_{\mathcal H}$ for a 3-sublattice
AFQ state can be calculated by performing a gradient expansion for the 3-site, 9-bond cluster 
shown in Fig.~\ref{fig:triangularlattice}, following the example of Eq.~(\ref{eq:gradexpansion}).
This gives
 \begin{align}
& \Delta\mathcal{H}_{\sf SU(2),clus} \approx 
  3(J_2-J_1) \left(  
  |{\bf d}_{\sf A}({\bf r},\tau)\cdot{\bf d}_{\sf B}({\bf r},\tau)|^2 \right. \nonumber \\
&\left. \qquad +|{\bf d}_{\sf B}({\bf r},\tau)\cdot{\bf d}_{\sf C}({\bf r},\tau)|^2+|{\bf d}_{\sf C}({\bf r},\tau)\cdot{\bf d}_{\sf A}({\bf r},\tau)|^2 \right) \nonumber \\
& \quad +\frac{3(J_2-J_1)a^2}{2} \sum_{\lambda ={\sf x,y}} \left[ |{\bf d}_{\sf A}\cdot\partial_ \lambda {\bf d}_{\sf B}|^2 
 +|{\bf d}_{\sf B}\cdot\partial_ \lambda {\bf d}_{\sf C}|^2 \right. \nonumber \\
&\left. \qquad \qquad \qquad \qquad \qquad \qquad +|{\bf d}_{\sf C}\cdot\partial_ \lambda {\bf d}_{\sf A}|^2 \right] 
 \nonumber \\
& \quad-\frac{3(J_2-J_1)a^2}{4}\sum_{\lambda ={\sf x,y}} \nonumber \\
& \qquad \left[ ({\bf d}_{\sf A}\cdot{\bf d}_{\sf B}) (\partial_ \lambda \bar{{\bf d}}_{\sf A}\cdot\partial_ \lambda \bar{{\bf d}}_{\sf B}) 
 + (\bar{{\bf d}}_{\sf A}.\bar{{\bf d}}_{\sf B}) (\partial_ \lambda{\bf d}_{\sf A}\cdot\partial_ \lambda{\bf d}_{\sf B})  \right. \nonumber \\
 &\qquad +({\bf d}_{\sf B}\cdot{\bf d}_{\sf C}) (\partial_ \lambda \bar{{\bf d}}_{\sf B}\cdot\partial_ \lambda \bar{{\bf d}}_{\sf C}) 
 + (\bar{{\bf d}}_{\sf B}.\bar{{\bf d}}_{\sf C}) (\partial_ \lambda{\bf d}_{\sf B}\cdot\partial_ \lambda{\bf d}_{\sf C}) \nonumber \\
 & \qquad +({\bf d}_{\sf C}\cdot{\bf d}_{\sf A}) (\partial_ \lambda \bar{{\bf d}}_{\sf C}\cdot\partial_ \lambda \bar{{\bf d}}_{\sf A}) 
 + (\bar{{\bf d}}_{\sf C}.\bar{{\bf d}}_{\sf A}) (\partial_ \lambda{\bf d}_{\sf C}\cdot\partial_ \lambda{\bf d}_{\sf A})
\left.  \right] ,
\label{eq:Hanisexpansion}
\end{align}
where the expansion has been truncated at second order in $a$.


Consider the product,
 \begin{align}
 {\bf d}_{\sf A}({\bf r},\tau)\cdot{\bf d}_{\sf B}({\bf r},\tau)\approx 
 \left(1,\bar{l}^{\sf z},l^{\sf y}  \right) \cdot {\bf U}^T{\bf U} \cdot
 \left(
\begin{array}{c}
l^{\sf z}  \\
1 \\
\bar{l}^{\sf x} 
\end{array}
\right),
\end{align}
where the matrices can be expressed as,
 \begin{align}
 {\bf U}^T{\bf U}=\left(
\begin{array}{ccc}
{\bf n}_{\sf A}^2 & {\bf n}_{\sf A}\cdot{\bf n}_{\sf B} & {\bf n}_{\sf C}\cdot{\bf n}_{\sf A}  \\
{\bf n}_{\sf A}\cdot{\bf n}_{\sf B} & {\bf n}_{\sf B}^2 & {\bf n}_{\sf B}\cdot{\bf n}_{\sf C} \\
 {\bf n}_{\sf C}\cdot{\bf n}_{\sf A} & {\bf n}_{\sf B}\cdot{\bf n}_{\sf C} & {\bf n}_{\sf C}^2
\end{array}
\right).
\end{align}
The ground state of the system involves purely real (or purely imaginary) ${\bf d}$ vectors, 
and therefore at low $T$ it is reasonable to approximate, 
 \begin{align}
{\bf n}_{\sf i}^2 \approx 1,
\end{align}
and,
 \begin{align}
{\bf n}_{\sf i}\cdot{\bf n}_{\sf j} \ll 1, \quad  {\sf i} \neq {\sf j}.
\end{align}
It follows that,
\begin{align}
{\bf n}_{\sf i}\cdot{\bf n}_{\sf j}&\approx -\bar{{\bf n}}_{\sf i}\cdot\bar{{\bf n}}_{\sf j}, 
\end{align}
and therefore,
 \begin{align}
 {\bf d}_{\sf A}({\bf r},\tau)\cdot{\bf d}_{\sf B}({\bf r},\tau) &\approx  l^{\sf z}+\bar{l}^{\sf z}+{\bf n}_{\sf A}\cdot{\bf n}_{\sf B} \nonumber \\
  {\bf d}_{\sf B}({\bf r},\tau)\cdot{\bf d}_{\sf C}({\bf r},\tau) &\approx  l^{\sf x}+\bar{l}^{\sf x}+{\bf n}_{\sf B}\cdot{\bf n}_{\sf C} \nonumber \\
   {\bf d}_{\sf C}({\bf r},\tau)\cdot{\bf d}_{\sf A}({\bf r},\tau) &\approx  l^{\sf y}+\bar{l}^{\sf y}+{\bf n}_{\sf C}\cdot{\bf n}_{\sf A}.
\end{align}
Using these approximations, the first term in Eq.~(\ref{eq:Hanisexpansion}) can be re-expressed as,
\begin{align}
  &|{\bf d}_{\sf A}({\bf r},\tau)\cdot{\bf d}_{\sf B}({\bf r},\tau)|^2+|{\bf d}_{\sf B}({\bf r},\tau)\cdot{\bf d}_{\sf C}({\bf r},\tau)|^2\nonumber \\
  &\qquad \qquad +|{\bf d}_{\sf C}({\bf r},\tau)\cdot{\bf d}_{\sf A}({\bf r},\tau)|^2 \approx \nonumber \\
  & (l^{\sf z}+\bar{l}^{\sf z})^2+(l^{\sf x}+\bar{l}^{\sf x})^2+(l^{\sf y}+\bar{l}^{\sf y})^2 \nonumber \\
  &\qquad \qquad +|{\bf n}_{\sf A}\cdot{\bf n}_{\sf B}|^2 +|{\bf n}_{\sf B}\cdot{\bf n}_{\sf C}|^2+|{\bf n}_{\sf C}\cdot{\bf n}_{\sf A}|^2.
\end{align}


Following the same procedure as in Section~\ref{sec:SU3} results in the Lagrangian,
\begin{align}
\mathcal{L}^{\sf SU(2)}_{\sf \triangle} \approx  & 
   \mathrm{Tr}[{\bf U}\dg\cdot\partial_\tau {\bf U}] 
   + 2\left[ ({\bf s}-\bar{{\bf s}}).{\bf l}_1 +i({\bf s}+\bar{{\bf s}}) {\bf l}_2 \right]  \nonumber \\
   &+12J_2  {\bf l}_1.{\bf l}_1 
  +12J_1  {\bf l}_2.{\bf l}_2  \nonumber \\
 & +3(J_2-J_1) (|{\bf n}_{\sf A}\cdot{\bf n}_{\sf B}|^2+|{\bf n}_{\sf B}\cdot{\bf n}_{\sf C}|^2+|{\bf n}_{\sf C}\cdot{\bf n}_{\sf A}|^2) \nonumber \\
& +\mathrm{gradient \ terms}.
\label{eq:Ltriangle}
  \end{align}
The canting fields ${\bf l}$ can once again be eliminated within a saddle-point approximation.
Performing the necessary functional derivative, and using Eq.~(\ref{eq:sfield}) to write 
the result in terms of ${\bf n}$, we find
\begin{align}
{\bf l}_1 &\approx -\frac{1}{12J_2}  \left(
\begin{array}{c}
\bar{{\bf n}}_{\sf B} . \partial_\tau {\bf n}_{\sf A} -{\bf n}_{\sf B}\cdot\partial_\tau \bar{{\bf n}}_{\sf A}  \\
\bar{{\bf n}}_{\sf C}\cdot\partial_ \tau {\bf n}_{\sf B} -{\bf n}_{\sf C}\cdot\partial_\tau \bar{{\bf n}}_{\sf B}  \\
\bar{{\bf n}}_{\sf A}\cdot\partial_ \tau {\bf n}_{\sf C} -{\bf n}_{\sf A}\cdot\partial_\tau \bar{{\bf n}}_{\sf C} 
\end{array}
\right) \nonumber \\
{\bf l}_2 &\approx -\frac{i}{12J_1} \left(
\begin{array}{c}
\bar{{\bf n}}_{\sf B}\cdot\partial_\tau {\bf n}_{\sf A} +{\bf n}_{\sf B}\cdot\partial_\tau \bar{{\bf n}}_{\sf A}  \\
\bar{{\bf n}}_{\sf C}\cdot\partial_ \tau {\bf n}_{\sf B} +{\bf n}_{\sf C}\cdot\partial_\tau \bar{{\bf n}}_{\sf B}  \\
\bar{{\bf n}}_{\sf A}\cdot\partial_ \tau {\bf n}_{\sf C} +{\bf n}_{\sf A}\cdot\partial_\tau \bar{{\bf n}}_{\sf C} 
\end{array}
\right).
\label{eq:mintermsofn}
\end{align}


These two canting fields correspond to physically distinct spin-- and quadrupole wave 
excitations.
These are no longer degenerate once the ${\sf SU(3)}$ symmetry is broken, 
and to parameterise them, we need to introduce two distinct susceptibilities, 
\begin{eqnarray}
\chi^{\sf Q}_\perp = \frac{2}{9J_1}  ,
\quad \quad
\chi^{\sf S}_\perp = \frac{2}{9J_2} ,
\end{eqnarray}
and two distinct director stiffnesses (which for this particular model, happen to be equal),
\begin{eqnarray}
\rho^{\sf Q}_{\sf d} = \rho^{\sf S}_{\sf d} = J_2 a^2.
\end{eqnarray}
It also proves convenient to reparamaterize the term in $\mathcal{L}^{\sf SU(2)}_{\sf \triangle}$ 
which breaks ${\sf SU(3)}$ symmetry in terms of a gap to spin wave excitations, i.e.
\begin{align}
\delta \mathcal{L}^{\sf SU(2)}_{\sf \triangle} = 
\frac{3}{8}\chi_\perp^{S} \Delta^2 
(|{\bf n}_{\sf A}\cdot{\bf n}_{\sf B}|^2+|{\bf n}_{\sf B}\cdot{\bf n}_{\sf C}|^2+|{\bf n}_{\sf C}\cdot{\bf n}_{\sf A}|^2)   
\end{align}
where
\begin{eqnarray}
\Delta = \sqrt{36J_2(J_2-J_1)}
\end{eqnarray}


Collecting these facts together, the action describing long-wavelength excitations of 
3-sublattice AFQ order is,
\begin{align}
& \mathcal{S}^{\sf SU(2)}_\triangle[
{\bf n}_{\sf A},
\bar{{\bf n}}_{\sf A},
{\bf n}_{\sf B},
\bar{{\bf n}}_{\sf B},
{\bf n}_{\sf C},
\bar{{\bf n}}_{\sf C}] 
\nonumber  \\
& =  \frac{1}{4\sqrt{3}} 
\int_0^\beta d\tau \int d^2r \left\{ 
 \frac{8}{3} \sum_{\sf i} \bar{{\bf n}}_{\sf i}\cdot\partial_\tau {\bf n}_{\sf i}  \right.
  \nonumber  \\
& + \chi_\perp^{\sf Q} \left[  (\bar{{\bf n}}_{\sf A}\cdot\partial_ \tau {\bf n}_{\sf B}+{\bf n}_{\sf A} 
   \cdot \partial_ \tau \bar{{\bf n}}_{\sf B})^2 \right. \nonumber  \\
&\qquad \left. +(\bar{{\bf n}}_{\sf B}\cdot\partial_ \tau {\bf n}_{\sf C}+{\bf n}_{\sf B}
   \cdot \partial_ \tau \bar{{\bf n}}_{\sf C})^2 \right. \nonumber\\
& \qquad \left. + (\bar{{\bf n}}_{\sf C}\cdot\partial_ \tau {\bf n}_{\sf A} + {\bf n}_{\sf C}
   \cdot \partial_ \tau \bar{{\bf n}}_{\sf A})^2 \right]   \nonumber \\  
& - \chi_\perp^{\sf S} \left[  (\bar{{\bf n}}_{\sf A}\cdot\partial_ \tau {\bf n}_{\sf B} - {\bf n}_{\sf A} 
   \cdot \partial_ \tau \bar{{\bf n}}_{\sf B})^2  \right. \nonumber  \\
& \qquad \left. +(\bar{{\bf n}}_{\sf B}\cdot\partial_ \tau {\bf n}_{\sf C}-{\bf n}_{\sf B}
   \cdot \partial_ \tau \bar{{\bf n}}_{\sf C})^2 \right.
\nonumber  \\
& \qquad \left.+ (\bar{{\bf n}}_{\sf C}\cdot\partial_ \tau {\bf n}_{\sf A}-{\bf n}_{\sf C}
   \cdot \partial_ \tau \bar{{\bf n}}_{\sf A})^2 \right] \nonumber  \\
& + \rho^{\sf Q}_{\sf d} \sum_{\lambda ={\sf x,y}} 
\left[ (\bar{{\bf n}}_{\sf A}\partial_ \lambda{\bf n}_{\sf B} 
+ {\bf n}_{\sf A}\partial_ \lambda\bar{{\bf n}}_{\sf B})^2 \right.
\nonumber  \\
& \qquad + (\bar{{\bf n}}_{\sf B}\partial_ \lambda{\bf n}_{\sf C} 
+ {\bf n}_{\sf B}\partial_ \lambda\bar{{\bf n}}_{\sf C})^2 
\nonumber  \\
& \left. \qquad + (\bar{{\bf n}}_{\sf C}\partial_ \lambda{\bf n}_{\sf A} 
+ {\bf n}_{\sf C}\partial_ \lambda\bar{{\bf n}}_{\sf A})^2 
\right] 
\nonumber \\
& - \rho^{\sf S}_{\sf d} \sum_{\lambda ={\sf x,y}} 
\left[ (\bar{{\bf n}}_{\sf A}\partial_ \lambda{\bf n}_{\sf B} 
- {\bf n}_{\sf A}\partial_ \lambda\bar{{\bf n}}_{\sf B})^2 \right.
\nonumber  \\
& \qquad + (\bar{{\bf n}}_{\sf B}\partial_ \lambda{\bf n}_{\sf C} 
- {\bf n}_{\sf B}\partial_ \lambda\bar{{\bf n}}_{\sf C})^2 
\nonumber  \\
& \left. \qquad + (\bar{{\bf n}}_{\sf C}\partial_ \lambda{\bf n}_{\sf A} 
- {\bf n}_{\sf C}\partial_ \lambda\bar{{\bf n}}_{\sf A})^2 
\right] 
\nonumber \\
&  \left. + \chi_\perp^{\sf S} \Delta^2 
(|{\bf n}_{\sf A}\cdot{\bf n}_{\sf B}|^2
+ |{\bf n}_{\sf B}\cdot{\bf n}_{\sf C}|^2
+ |{\bf n}_{\sf C}\cdot{\bf n}_{\sf A}|^2) \right\}
\label{eq:Striangle}
\end{align}
where the relevant parameters for the microscopic model 
${\mathcal H}_\triangle^{\sf BBQ}$ [Eq.~(\ref{eq:H-BBQ-triangle})]
are given in Table~\ref{table:3-sublattice-dictionary}, 
and the partition function is defined as in Eq.~(\ref{eq:ZSU3}).


\begin{figure*}[ht]
\centering
\includegraphics[width=0.8\textwidth]{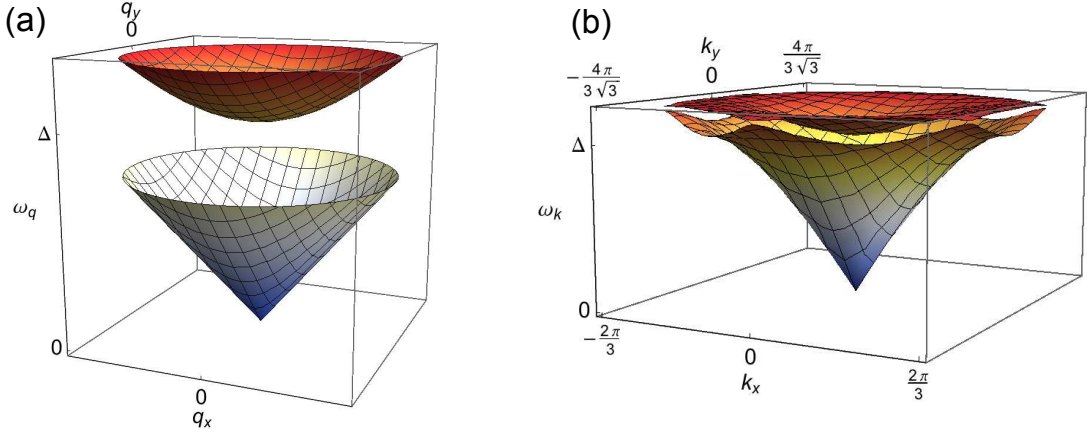}
\caption{\footnotesize{(Color online). 
Dispersion of magnetic excitations in 
a 3-sublattice antiferroquadrupolar (AFQ) spin-nematic state 
on a triangular lattice.
(a) prediction of the continuum field theory 
$\mathcal{S}^{\sf SU(2)}_\triangle[\boldsymbol{\phi}]$ [Eq.~(\ref{eq:linaction})], with dispersion given by $\omega_{\bf q}^{\sf Q}$~[Eq.~(\ref{eq:omega-})] and $\omega_{\bf q}^{\sf S}$~[Eq.~(\ref{eq:omega+})].  
(b) prediction of the microscopic model 
${\mathcal H}^{\sf BBQ}_\triangle$~[Eq.~(\ref{eq:H-BBQ-triangle})] in the magnetic Brillouin zone (mbz) [see Fig.~\ref{fig:mbz}], for parameters $J_1=1$ and $J_2=1.22$.
The dispersion is given by $\omega^\pm_{\bf q}$~[Eq.~(\ref{eq:3subAFQflavourwave})].
In both cases a three-fold degenerate branch of gapless, quadrupole-wave excitations, 
are centered on the ordering vector ${\bf k}={\bf k}_{\sf K}$ [ie. {\bf q}=0].
These are the Goldstone modes of the AFQ order.  
They are accompanied by a three-fold degenerate branch of gapped, spin-wave 
excitations.
Approaching the centre of the mbz,  ${\bf q} \to 0$, 
the continuum theory and the lattice theory match exactly.
}}
\label{fig:dispersionwithgap}
\end{figure*}


Eq.~(\ref{eq:Uvsn}) can be used to re-express this action in terms of the unitary matrix field, ${\bf U}({\bf r},\tau)$, as,
\begin{align}
&\mathcal{S}^{\sf SU(2)}_\triangle[{\bf U}] =   \frac{1}{8\sqrt{3}a^2} \int_0^\beta d\tau \int d^2r
\left\{ \frac{16}{3}\mathrm{Tr}[{\bf U}\dg\cdot\partial_\tau {\bf U}] \right.\nonumber \\
& + \chi^{\sf Q}_\perp
\mathrm{Tr}\left( {\bf U}\dg\cdot\partial_\tau {\bf U} + {\bf U}^\mathrm{T} \cdot\partial_\tau \bar{\bf U} \right)\dg 
\left( {\bf U}\dg\cdot\partial_\tau {\bf U} + {\bf U}^\mathrm{T} \cdot\partial_\tau \bar{\bf U} \right)    \nonumber \\
& + \chi^{\sf S}_\perp
\left[ \mathrm{Tr}\left( {\bf U}\dg\cdot\partial_\tau {\bf U} - {\bf U}^\mathrm{T} \cdot\partial_\tau \bar{\bf U} \right)\dg 
\left( {\bf U}\dg\cdot\partial_\tau {\bf U} - {\bf U}^\mathrm{T} \cdot\partial_\tau \bar{\bf U} \right)  \right. \nonumber \\
&\left.\qquad  -4\sum_m \left| [{\bf U}\dg\cdot\partial_ \tau {\bf U}]_{mm} \right|^2  \right] \nonumber \\
& + \rho_{\sf d}^{\sf Q} 
\mathrm{Tr}\left( {\bf U}\dg\cdot\partial_\lambda {\bf U} + {\bf U}^\mathrm{T} \cdot\partial_ \lambda \bar{\bf U} \right)\dg 
\left( {\bf U}\dg\cdot\partial_ \lambda {\bf U} + {\bf U}^\mathrm{T} \cdot\partial_ \lambda \bar{\bf U} \right)    \nonumber \\
& + \rho_{\sf d}^{\sf S}
\left[ \mathrm{Tr}\left( {\bf U}\dg\cdot\partial_ \lambda {\bf U} - {\bf U}^\mathrm{T} \cdot\partial_ \lambda \bar{\bf U} \right)\dg 
\left( {\bf U}\dg\cdot\partial_\lambda {\bf U} - {\bf U}^\mathrm{T} \cdot\partial_ \lambda \bar{\bf U} \right)  \right. \nonumber \\
&\left.\qquad  -4\sum_m \left| [{\bf U}\dg\cdot\partial_ \tau {\bf U}]_{mm} \right|^2  \right] \nonumber \\
& \left. + \chi^{\sf S}_\perp \Delta^2 
\left[ 3-\sum_i \left| [{\bf U}^\mathrm{T}.{\bf U}]_{mm} \right|^2 \right] 
\right\}.
\label{eq:S-of-U-triangle}
\end{align}
This reduces to Eq.~(\ref{eq:SSU3Umat}) when $\chi_\perp^{\sf Q}=\chi_\perp^{\sf S}$ and $\Delta=0$ (ie. $J_1=J_2$), as required.


\begin{table}
\begin{center}
  \begin{tabular}{| c | c | }
    \hline
\multirow{2}{*}{ $\mathcal{S}^{\sf SU(2)}_\triangle[ {\bf U} ]$ }  &  
\multirow{2}{*}{ ${\mathcal H}^{\sf BBQ}_\triangle$ }    \\ 
&   \\ \hline
\multirow{2}{*}{$\chi_\perp^{\sf Q}$}  &  
\multirow{2}{*}{$2/(9J_1)$}    \\ 
&   \\ \hline
\multirow{2}{*}{$\chi_\perp^{\sf S}$}  &  
\multirow{2}{*}{$2/(9J_2)$}    \\ 
&   \\ \hline
\multirow{2}{*}{$\rho^{\sf Q}_{\sf d}$}  &  
\multirow{2}{*}{$J_2 a^2$}    \\ 
&   \\ \hline
\multirow{2}{*}{$\rho^{\sf S}_{\sf d}$}  &  
\multirow{2}{*}{$J_2 a^2$}    \\ 
&   \\ \hline
\multirow{2}{*}{$\Delta$}  &  
\multirow{2}{*}{$\sqrt{36J_2(J_2-J_1)}$}    \\ 
&   \\ \hline
    \end{tabular}
\end{center} 
\caption{
\footnotesize{
Dictionary for translating between the parameters of the 
continuum field theory for 3-sublattice AFQ order, 
$\mathcal{S}^{\sf SU(2)}_\triangle[{\bf U}]$ [Eq.~(\ref{eq:S-of-U-triangle})],
and the parameters of the relevant microscopic model 
${\mathcal H}^{\sf BBQ}_\triangle$~[Eq.~(\ref{eq:H-BBQ-triangle})],
in the vicinity of the ${\sf SU(3)}$ point $J_1=J_2$.
}}
\label{table:3-sublattice-dictionary}
\end{table}


\subsubsection{Linearising the order parameter fields}


The physical content of the action $\mathcal{S}^{\sf SU(2)}_\triangle[{\bf n}]$ [Eq.~(\ref{eq:Striangle})], 
becomes clear on linearisation of the fields. 
Once again, we use Eq.~(\ref{eq:linearisation}) to expand small fluctuations about the 
ground state in terms of $\boldsymbol{\phi}$.
This leads to the action 
\begin{align}
&\mathcal{S}^{\sf SU(2)}_\triangle[\boldsymbol{\phi}] \approx \frac{1}{\sqrt{3}a^2} \int_0^\beta d\tau  \int d^2r  \nonumber \\
&\qquad \sum_{p=1\dots 3} \left[ \chi_\perp^{\sf Q}(\partial_\tau\phi_p)^2 + \rho_{\sf d}^{\sf Q} \sum_{\lambda ={\sf x,y}}(\partial_\lambda\phi_p)^2 \right]\nonumber \\
& \qquad +\sum_{p=4\dots 6} \left[ \chi_\perp^{\sf S}(\partial_\tau\phi_p)^2 + \rho_{\sf d}^{\sf S} \sum_{\lambda ={\sf x,y}}(\partial_\lambda\phi_p)^2 +\chi_\perp^{\sf S} \Delta^2 \phi_p^2 \right]
\label{eq:linaction}
\end{align}
We immediately see that there are three gapless, quadrupole-wave modes, 
$\phi_1$, $\phi_2$ and $\phi_3$, with dispersion,
\begin{align}
\omega_{\bf q}^{\sf Q} \approx v_{\sf Q} |{\bf q}|, \quad 
v_{\sf Q}=\sqrt{\frac{\rho_{\sf d}^{\sf Q}}{\chi^{\sf Q}_\perp}} = 3\sqrt{\frac{J_1J_2}{2}}a,
\label{eq:omega-}
\end{align}
and three gapped, spin-wave modes $\phi_4$, $\phi_5$ and $\phi_6$, with dispersion
\begin{align}
\omega_{\bf q}^{\sf S} \approx \sqrt{\Delta^2+v_{\sf S}^2{\bf q}^2}, \quad
 v_{\sf S}=\sqrt{\frac{\rho_{\sf d}^{\sf S}}{\chi^{\sf S}_\perp}}= 3\frac{J_2a}{\sqrt{2}}.
\label{eq:omega+}
\end{align}
These are shown in Fig.~\ref{fig:dispersionwithgap}. 
The Goldstone modes correspond to real rotations of the order parameter fields, while the 
gapped modes (gap $\Delta$) correspond to rotations into complex space.


The microscopic `flavour-wave' theory developed by Tsunetsugu and Arikawa [\onlinecite{tsunetsugu06,tsunetsugu07}] predicts a dispersion,
\begin{align}
\omega^{\pm}_{\bf k} =3J_2\sqrt{\left(1\pm |\gamma_{\bf k}| \right) \left(1\pm \left( 1-\frac{2J_1}{J_2} \right) |\gamma_{\bf k}| \right) },
\label{eq:3subAFQflavourwave}
\end{align}
where $\gamma_{\bf k}$ is given by Eq.~(\ref{eq:gamma}). 
This is shown in Fig.~\ref{fig:dispersionSU3}. 
In the long wavelength limit, and for small \mbox{$J_2-J_1$}, the dispersion reduces 
to Eq.~(\ref{eq:omega-}) and Eq.~(\ref{eq:omega+}). 


We re-emphasise that the validity of the continuum theory breaks down approaching 
the FM phase for \mbox{$\theta \to \pi/2$} ($J_1 \to 0$, $J_2 > 0$).
Crossing the AFQ phase, there is a progressive reduction in the area of reciprocal 
space over which the quadrupole-wave dispersion, $\omega^-_{\bf k}$, is linear. 
This is also a feature of the lattice theory ---exactly at the phase boundary with the 
ferromagnet ($J_1=0$, $J_2 > 0$) the dispersion, $\omega^{\pm}_{\bf k}$ 
[Eq.~(\ref{eq:3subAFQflavourwave})], becomes quadratic even for $|{\bf k}| \to 0$.  
This signals that it is no longer appropriate to describe the system in terms of the 
quantum non-linear sigma model, $\mathcal{S}^{\sf SU(2)}_\triangle[{\bf U}]$, [Eq.~(\ref{eq:S-of-U-triangle})].


\subsection{The low temperature, low energy limit}


For temperature and energy scales lower than the spin wave gap, $\Delta$, the high-energy, 
spin-wave modes can be neglected. 
This considerably simplifies the action, $\mathcal{S}^{\sf SU(2)}_\triangle[{\bf n}]$ [Eq.~(\ref{eq:Striangle})], 
and is a useful approximation when considering low temperature thermodynamic properties.


Neglection of the spin-wave modes is equivalent to making the assumption that the fields 
${\bf n}_{\sf i}$, are real. 
The simplified action is then given by,
\begin{align}
&\mathcal{S}^{\sf SO(3)}_{\sf \triangle}[{\bf n}_{\sf A},{\bf n}_{\sf B},{\bf n}_{\sf C}] \approx \frac{1}{2\sqrt{3}a^2} \int_0^\beta d\tau  \int d^2r  \nonumber \\
& \qquad \{ \chi_\perp^{\sf Q}\left[(\partial_\tau{\bf n}_{\sf A})^2+(\partial_\tau{\bf n}_{\sf B})^2+(\partial_\tau{\bf n}_{\sf C})^2\right] \nonumber \\
& \qquad  + \rho_{\sf d}^{\sf Q}  \sum_{\lambda ={\sf x,y}}\left[(\partial_\lambda{\bf n}_{\sf A})^2+(\partial_ \lambda{\bf n}_{\sf B})^2+(\partial_ \lambda{\bf n}_{\sf 
C})^2\right] \},
\label{eq:SO(3)action}
\end{align}
with canting field,
\begin{align}
{\bf l}\approx i {\bf l}_2 &\approx \frac{3}{4} \chi_\perp^{\sf Q} \left(
\begin{array}{c}
{\bf n}_{\sf B}\cdot\partial_\tau {\bf n}_{\sf A}  \\
{\bf n}_{\sf C}\cdot\partial_ \tau {\bf n}_{\sf B}   \\
{\bf n}_{\sf A}\cdot\partial_ \tau {\bf n}_{\sf C} 
\end{array}
\right),
\end{align}
and the partition function is,
\begin{align}
\mathcal{Z}^{\sf SO(3)}_{\sf \triangle} \propto & \int \prod_{{\sf i}\neq {\sf j}}
\mathcal{D}{\bf n}_{\sf i} 
\ \delta({\bf n}_{\sf i}^2-1) 
\ \delta({\bf n}_{\sf i}\cdot{\bf n}_{\sf j}) 
\ e^{-\mathcal{S}^{\sf SU(2)}_\triangle[{\bf n}_{\sf A},{\bf n}_{\sf B},{\bf n}_{\sf C}]}.
\end{align}


This is an $SO(3)$ symmetric non-linear sigma model\cite{azaria93}, 
a fact which is clearer if the action is written in matrix form,
\begin{align}
&\mathcal{S}^{\sf SO(3)}_{\sf \triangle}[ {\bf R}] \approx \frac{1}{2\sqrt{3}a^2} \int_0^\beta d\tau  \int d^2r  \nonumber \\
& \qquad\{ \chi_\perp^{\sf Q} \mathrm{Tr} \left[  \partial_\tau {\bf R}^\mathrm{T}\cdot\partial_\tau {\bf R} \right]
 + \rho_{\sf d}^{\sf Q} \sum_{\lambda={\sf x,y}} \mathrm{Tr}  \left[  \partial_ \lambda {\bf R}^\mathrm{T}\cdot\partial_ \lambda {\bf R} \right] \}, 
\end{align}
where ${\bf R}$ is a real-valued rotation matrix given by,
\begin{align}
 {\bf R}({\bf r},\tau)= \left(
\begin{array}{ccc}
n_{\sf A}^{\sf x}({\bf r},\tau) & n_{\sf B}^{\sf x}({\bf r},\tau) & n_{\sf C}^{\sf x}({\bf r},\tau)  \\
n_{\sf A}^{\sf y}({\bf r},\tau) & n_{\sf B}^{\sf y}({\bf r},\tau)  & n_{\sf C}^{\sf y}({\bf r},\tau) \\
n_{\sf A}^{\sf z}({\bf r},\tau) & n_{\sf B}^{\sf z}({\bf r},\tau) & n_{\sf C}^{\sf z}({\bf r},\tau)
\end{array}
\right).
\label{eq:Rvsn}
\end{align}


The simplified action, Eq.~(\ref{eq:SO(3)action}), describes the 3 quadrupole-wave modes shown in 
Fig.~\ref{fig:dispersionwithgap} but ignores the 3 spin-wave modes which dominate experimental
responses at higher energy.  


\subsection{Comparision with other forms of magnetic order}


It is interesting to compare the continuum theory of long-wavelength excitations in a 
3-sublattice AFQ state, $\mathcal{S}^{\sf SU(2)}_\triangle[{\bf U}]$ [Eq.~(\ref{eq:S-of-U-triangle})], 
with sigma-model approaches to other forms of magnetic order.
Perhaps the most widely known example is the sigma-model treatment 
of the collinear antiferromagnet (AFM) [\onlinecite{fradkin,auerbach,chakravarty89,allen97}].
The collinear nature of this state means that it does not break the full {\sf SU(2)} 
spin-rotation symmetry, but instead {\sf SU(2)}/{\sf U(1)}. 
As a consequence the resulting sigma model describes only two, degenerate, 
linearly-dispersing Goldstone modes, both with the character of spin-wave excitations.
The only gapped excitation possible at long wavelength is a 
longitudinal fluctuation of the order parameter, explicitly absent from the sigma model.
Collinearity also imposes constrains on the interactions which can arise between different 
spin excitations, restricting these to vertices involving an even number of excitations.


The key features of the continuum theory of \mbox{3-sublattice} AFQ order, 
$\mathcal{S}^{\sf SU(2)}_\triangle[{\bf U}]$ [Eq.~(\ref{eq:S-of-U-triangle})], are three 
degenerate, linearly-dispersing ``quadrupole-wave'' modes associated with the breaking of 
spin-rotation symmetry, and three degenerate, gapped ``spin-wave'' modes, 
associated with dipolar excitations of the underlying quadrupolar order.
The two actions therefore differ in both the number and the character of the modes
they describe.
It is also worth noting that, the structure of the interactions between these excitations 
(not described in this article) is profoundly different, and includes vertices with an odd
number of excitations.
This topic will be explored further elsewhere~\cite{smerald-unpub}.


The action $\mathcal{S}^{\sf SU(2)}_\triangle[{\bf U}]$ [Eq.~(\ref{eq:S-of-U-triangle})] finds more 
parallels with non-collinear magnetic ordering.
A good example of this is the $120^\circ$ state on the triangular lattice\cite{dombre89,apel92}.
This fully breaks the {\sf SU(2)} symmetry, and therefore has three, linearly-dispersing 
Goldstone modes, all with the character of spin waves.
Interactions between odd numbers of spin excitations are also now permitted by symmetry.
However, as with the collinear antiferromagnet, the $120^\circ$ state has no low-energy gapped 
modes at long-wavelength.
Also, the coplanar nature of this state means that the spin stiffness' associated with the three 
Goldstone modes are not all equal, and only two of the three Goldstone modes are degenerate.


Finally it is interesting to compare $\mathcal{S}^{\sf SU(2)}_\triangle[{\bf U}]$ [Eq.~(\ref{eq:S-of-U-triangle})] 
with field theories describing FQ order \cite{ivanov03,ivanov07,ivanov08,baryakhtar13}.
As with the collinear AFM, FQ states have only two Goldstone modes.
These are degenerate, linearly dispersing, and have the character of quadrupole waves at 
long wavelength.
Only interactions between even numbers of spin excitations are permitted by symmetry.
Both of these points clearly distinguish the present theory of AFQ order from the 
earlier work on FQ order.


In fact the theory derived in Ref.~[\onlinecite{ivanov03}] has the same action as the collinear 
AFM, albeit with a different physical interpretation.
However, in reducing the action to this form, imaginary fluctuations of the director {\bf d}
have been explicitly integrated out, eliminating much of the information concerning excitations 
with ``spin-wave'' character.
An important feature of the ${\sf SU(3)}$-derived approach developed in this article is
its ability to describe gapped excitations with dipolar character, such as the ``spin-wave'' 
modes of AFQ order, which {\it cannot} be accessed in the ${\sf SO(3)}$ approach of 
Ref.~[\onlinecite{ivanov03}].
Such modes are particularly interesting since they will be the easiest to observe in, e.g., 
inelastic neutron scattering. 


\subsection{Machinery for calculating correlation functions}
\label{sec:3-sublattice-spin-toolbox}


In order to make predictions for inelastic neutron scattering and for the dynamical quadrupole susceptibility, 
it is necessary to translate the continuum field theory,
$\mathcal{S}^{\sf SU(2)}_\triangle[{\bf U}]$~[Eq.~(\ref{eq:S-of-U-triangle})]
--- which is written in terms of rotations of directors --- 
back into the language of spins and quadrupoles.
Following 
Eq.~(\ref{eq:directors-as-U-times-l}) 
and 
Eq.~(\ref{eq:linearized-U}), 
the directors on the three sublattices can be approximated as,
\begin{align}
{\bf d}_{\sf A} & \approx
\left(
\begin{array}{c}
1 \\
-\phi_1+i\phi_4 +\bar{l}^{\sf z}  \\
\phi_2+i\phi_5+l^{\sf y}
\end{array}
\right)
 \nonumber \\
{\bf d}_{\sf B} & \approx
\left(
\begin{array}{c}
\phi_1+i\phi_4 +l^{\sf z}  \\
1 \\
-\phi_3+i\phi_6+\bar{l}^{\sf x}
\end{array}
\right)
\nonumber \\
{\bf d}_{\sf C} & \approx
\left(
\begin{array}{c}
-\phi_2+i\phi_5 +\bar{l}^{\sf y} \\
\phi_3+i\phi_6+l^{\sf x}  \\
1
\end{array}
\right),
\end{align}
with the canting fields,
\begin{align}
{\bf l}_1\approx -\frac{3}{4}\chi_\perp^{\sf S}  \left(
\begin{array}{c}
\partial_t \phi_4  \\
\partial_t \phi_6   \\
\partial_t \phi_5 
\end{array}
\right), \quad
{\bf l}_2 \approx \frac{3}{4}\chi_\perp^{\sf Q} \left(
\begin{array}{c}
\partial_t \phi_1  \\
\partial_t \phi_3 \\
\partial_t \phi_2
\end{array}
\right),
\label{eq:lintermsofphi}
\end{align}
where the real time $t=-i\tau$ has been used. 
It follows that the ${\bf d}$ vectors are,
\begin{align}
{\bf d}_{\sf A} & \approx \left(
\begin{array}{c}
1 \\
-\phi_1+i\phi_4 - \frac{3}{4} \chi_\perp^{\sf S} \partial_t \phi_4 - i\frac{3}{4} \chi_\perp^{\sf Q} \partial_t \phi_1 \\
\phi_2+i\phi_5- \frac{3}{4} \chi_\perp^{\sf S} \partial_t \phi_5 + i\frac{3}{4} \chi_\perp^{\sf Q} \partial_t \phi_2
\end{array}
\right) \nonumber \\
{\bf d}_{\sf B} & 
\approx \left(
\begin{array}{c}
\phi_1+i\phi_4 - \frac{3}{4} \chi_\perp^{\sf S} \partial_t \phi_4 + i\frac{3}{4} \chi_\perp^{\sf Q} \partial_t \phi_1 \\
1 \\
-\phi_3+i\phi_6- \frac{3}{4} \chi_\perp^{\sf S} \partial_t \phi_6 - i\frac{3}{4} \chi_\perp^{\sf Q} \partial_t \phi_3
\end{array}
\right) \nonumber \\
{\bf d}_{\sf C} &
\approx \left(
\begin{array}{c}
-\phi_2+i\phi_5 - \frac{3}{4} \chi_\perp^{\sf S} \partial_t \phi_5 - i\frac{3}{4} \chi_\perp^{\sf Q} \partial_t \phi_2 \\
\phi_3+i\phi_6- \frac{3}{4} \chi_\perp^{\sf S} \partial_t \phi_6 + i\frac{3}{4} \chi_\perp^{\sf Q} \partial_t \phi_3 \\
1
\end{array}
\right).
\end{align}


Substituting these expressions into Eq.~(\ref{eq:SQduvmatrix}) leads to the fluctuating dipole moments,
\begin{align}
{\bf S}_{\sf A} &\approx    \left(
\begin{array}{c}
0    \\
-2 \phi_5 -\frac{3}{2} \chi_\perp^{\sf Q}\partial_t \phi_2  \\
2\phi_4-\frac{3}{2} \chi_\perp^{\sf Q}\partial_t\phi_1
\end{array}
\right)
 \nonumber \\
{\bf S}_{\sf B} &\approx   \left(
\begin{array}{c}
2\phi_6-\frac{3}{2} \chi_\perp^{\sf Q} \partial_t \phi_3    \\
0  \\
-2\phi_4-\frac{3}{2} \chi_\perp^{\sf Q} \partial_t\phi_1
\end{array}
\right)
 \nonumber \\
{\bf S}_{\sf C} &\approx    \left(
\begin{array}{c}
-2 \phi_6-\frac{3}{2} \chi_\perp^{\sf Q}\partial_t \phi_3    \\
2 \phi_5-\frac{3}{2} \chi_\perp^{\sf Q}\partial_t \phi_2  \\
0
\end{array}
\right),
\label{eq:isotropicspins}
\end{align}
where terms linear in the $\boldsymbol{\phi}$ fields have been retained.
Eq.~(\ref{eq:isotropicspins}) provides the starting point for the theory of 
inelastic neutron scattering developed in Section~\ref{sec:neutrons} of this paper.


The quadrupole moments are given by,
\begin{align}
{\bf Q}_{\sf A} &\approx    \left(
\begin{array}{c}
-1    \\
\frac{1}{\sqrt{3}} \\
-2 \phi_1 -\frac{3}{2} \chi_\perp^{\sf S}\partial_t \phi_4  \\
0 \\
-2\phi_2+\frac{3}{2} \chi_\perp^{\sf S}\partial_t\phi_5
\end{array}
\right)
 \nonumber \\
{\bf Q}_{\sf B} &\approx    \left(
\begin{array}{c}
1    \\
\frac{1}{\sqrt{3}} \\
2 \phi_1 -\frac{3}{2} \chi_\perp^{\sf S}\partial_t \phi_4  \\
-2\phi_3-\frac{3}{2} \chi_\perp^{\sf S}\partial_t\phi_6 \\
0
\end{array}
\right)
 \nonumber \\
{\bf Q}_{\sf C} &\approx    \left(
\begin{array}{c}
0    \\
-\frac{2}{\sqrt{3}} \\
0 \\
2 \phi_3 -\frac{3}{2} \chi_\perp^{\sf S}\partial_t \phi_6  \\
2\phi_2+\frac{3}{2} \chi_\perp^{\sf S}\partial_t\phi_5
\end{array}
\right).
\label{eq:Qmoments}
\end{align}
The supplemental material contains animations showing the nature of the quadrupole-wave\cite{AFQ-Qwave-animation} and spin-wave\cite{AFQ-Swave-animation} excitations.


\section{Predictions for inelastic neutron scattering}
\label{sec:neutrons}


\subsection{General considerations : waves in the unseen}


Since each ``spin'' in a quantum magnet posseses a magnetic dipole, 
conventional {\it dipolar} magnetic order gives rise to a static internal magnetic field.
Neutrons, which also posses a dipole moment, diffract from this static 
field to give magnetic Bragg peaks.
As in conventional crystallography, the form of magnetic order present 
is encoded in the wave number and intensity of these magnetic Bragg peaks.
However, since spin-nematic order corresponds to a {\it quadrupolar} 
order of spins, it does {\it not} break time-reversal symmetry
and {\it cannot} give rise to static magnetic fields~\cite{penc11,andreev84}. 
For this reason, it does not manifest itself through magnetic 
Bragg peaks in elastic neutron scattering. 


An elegant solution to this problem, {\it in the presence of an anisotropy} that breaks ${\sf SU(2)}$ symmetry, was proposed by Barzykin and Gorkov~\cite{barzykin93}, 
who suggested using an external magnetic field to break time-reversal symmetry.
In the presence of magnetic anisotropy, applying a uniform magnetic field to an 
AFQ state induces a small, staggered, dipole moment which can, in principle, 
be observed in elastic neutron scattering.
Resonant magnetic X-ray scattering, which {\it is} sensitive to quadrupole moments of spins, 
has also been used to identify AFQ order in the rare-earth magnet UPd$_3$~\cite{mcmorrow01,walker06,walker08}
However a very direct and appealing route to identifying spin-nematic order, even in the absence of magnetic anisotropy,
would be to map out its magnetic excitations using inelastic neutron scattering.


Since spin-nematic order breaks spin-rotation symmetry it {\it must} possess Goldstone 
modes.   
The long-wavelength excitations are generated by real ${\sf SU(2)}$ rotations of the 
underlying quadrupolar order parameter, and so can best be thought of as ``quadrupole 
waves''.
Quadrupole waves possess a small fluctuating dipole moment, and will reveal themselves 
as linearly-dispersing excitations --- visible waves in the unseen spin-nematic order.
As we will see in what follows, the size of this dipole moment is directly proportional 
to the speed at which the quadrupoles rotate, and so the intensity of scattering 
from a quadrupole wave vanishes linearly with its energy.


However, precisely because the building blocks of spin-nematic order are quadrupole 
moments of spins, these Goldstone modes
do {\it not} exhaust the possible excitations of a spin-nematic state.
Neutrons can also drive transitions between different triplet states, which mix a strong spin-dipole
into the underlying quadrupole moment.   
In AFQ spin-nematic states, this leads to a second, distinct, type of long-wavelength excitation,
with a gapped spectrum and a pronounced intensity in inelastic neutron scattering.
Identifying this gapped excitation in experiment, together with the appropriate set of gapless
Goldstone modes, would provide strong evidence for the existence of spin-nematic order.


In Section~\ref{sec:3-sublattice-AFQ} of this paper 
we have developed the tools needed to make distinctive, quantitative predictions for
{\it both} types of excitation of a spin-nematic  --- a continuum field-theory of the excitations 
of AFQ order based on the symmetries of the underlying order parameter.
This ${\sf SU(3)}$ ``sigma-model'' approach offers a {\it quantitative} description of 
excitations --- in terms of the minimum set of physically meaningful parameters --- 
{\it without} the need to specify a microscopic model.


In what follows we use this continuum theory to make predictions for inelastic neutron scattering
carried out on a 3-sublattice AFQ state.
These predictions are {\it exact} at long wavelength, and fully constrain 
the symmetries broken by the AFQ state.
We make explicit comparison with the predictions of a microscopic, spin-1 lattice 
model that realises the same ordered state.
In order to keep the discussion reasonably self-contained, key results from 
Section~\ref{sec:3-sublattice-AFQ} are quoted in the text.


\subsection{Sum rules and correlation functions}


Inelastic neutron scattering measures the imaginary part of the dynamical spin susceptibility, 
\begin{align}
&\Im m \{\chi_{\sf S}^{\alpha\beta}({\bf k}, \omega)\} \nonumber \\
& = (g\mu_{\sf B})^2  \Im m \{ 
i \int_0^\infty dt 
e^{i \omega t} 
\langle \delta S^\alpha({\bf k}, t) \delta S^\beta(-{\bf k}, 0) \rangle
\}  \nonumber \\
\label{eq:im-chi}
\end{align}
where $\alpha, \beta = x,y,z$ label spin components.  
In the case of the 3-sublattice AFQ state described in Section~\ref{sec:3-sublattice-AFQ}, 
this tensor is diagonal, and fluctuations are isotropic in spin space, i.e.
\begin{align}
&\Im m \{\chi_{\sf S}^{xx}({\bf k}, \omega)\} = \Im m \{ \chi_{\sf S}^{yy}({\bf k}, \omega) \} = \Im m \{ \chi_{\sf S}^{zz}({\bf k}, \omega) \}.
\nonumber
\end{align}
%


An important check on any calculation of the dynamical susceptibility is that it obeys the relevant sum rules.
For any theory with {\sf SU(2)} spin symmetry, as is the case for the 3-sublattice AFQ state, it is required that,
 \begin{align}
\lim_{{\bf q}\to 0}\int d\omega \
e^{i \omega t}  \omega  \chi_{\sf S}^{\alpha\beta}({\bf k},\omega) =0.
\label{eq:sumrule}
\end{align}
This says that at ${\bf k}=0$, the dynamical susceptibility must vanish for all $\omega\neq 0$.
The sum rule is related to a Ward-Takahashi identity, and thus holds at {\it each} order in perturbation theory.
For single particle excitations it is sufficient to consider the non-interacting theory described by $\mathcal{S}^{\sf SU(2)}_\triangle[\boldsymbol{\phi}]$ [Eq.~(\ref{eq:linaction})].
However, in order to understand the 2-particle continuum it is necessary to take three and four field interactions into account and form a Dyson equation for the self energy.
Since this is an involved process, we postpone discussion until a future publication\cite{smerald-unpub}.
We note that the linear flavour wave analysis of Tsunetsugu and Arikawa [\onlinecite{tsunetsugu06}] obeys the sum rule, Eq.~(\ref{eq:sumrule}), at leading order, but has finite weight at ${\bf k}=0$ and $\omega\neq 0$ arising from the 2-particle continuum.


In Section~\ref{sec:quad-suscep} we also consider the dynamical quadrupole susceptibility.
This is given by,
\begin{align}
&\Im m \{\chi_{\sf Q}^{\alpha\beta\gamma\delta}({\bf k}, \omega)\} \nonumber \\
& = (g\mu_{\sf B})^4  \Im m \{ 
i \int_0^\infty dt 
e^{i \omega t} 
\langle \delta Q^{\alpha\beta}({\bf k}, t)  \delta Q^{\gamma\delta}(-{\bf k}, 0) \rangle
\}. \nonumber \\
\label{eq:im-chi}
\end{align}
In the general case of {\sf SU(2)} spin symmetry, there is no analogous sum rule to Eq.~(\ref{eq:sumrule}), and one expects to find finite weight at ${\bf k}=0$ and $\omega\neq 0$.
However, exactly at the {\sf SU(3)} point the expanded symmetry leads to the quadrupolar sum rule,
 \begin{align}
\lim_{{\bf k}\to 0}\int d\omega \
e^{i \omega t}  \omega  \chi_{\sf Q}^{\alpha\beta}({\bf k},\omega) = 0 \quad [{\sf SU(3)} \ \mathrm{point}].
\label{eq:sumrule}
\end{align}


\subsection{Neutron scattering in a 3-sublattice AFQ}
\label{sec:chiS}


\subsubsection{Spin excitations in a 3-sublattice AFQ state}
\label{sec:spin}


Predictions for inelastic neutron scattering from a 3-sublattice AFQ state have previously been 
published by Tsunetsugu and Arikawa~\cite{tsunetsugu06,tsunetsugu07}, based on flavour-wave calculations for 
the spin-1 bilinear-biquadratic (BBQ) model on the triangular lattice 
${\mathcal H}_\triangle^{\sf BBQ}$ [Eq.~(\ref{eq:H-BBQ-triangle})].
In what follows we show how the universal, long-wavelength features of these results
are completely described by the field theory developed in Section~\ref{sec:3-sublattice-AFQ} of this paper.
The tools needed to calculate $\Im m \{\chi_{\sf S}^{\alpha\beta}({\bf k}, \omega)\}$ --- namely a theory 
of long-wavelength spin excitations in a spin-nematic state --- were developed in 
Section~\ref{sec:3-sublattice-AFQ} of this paper.
Here we briefly reprise the most relevant results.


Small fluctuations about the 3-sublattice AFQ ordered state can be described by 
the linearized action, $\mathcal{S}^{\sf SU(2)}_\triangle[\boldsymbol{\phi}]$ [Eq.~(\ref{eq:linaction})], viz~:
\begin{align}
&\mathcal{S}^{\sf SU(2)}_\triangle[\boldsymbol{\phi}] \approx \frac{1}{\sqrt{3}a^2} \int_0^\beta d\tau  \int d^2r  \nonumber \\
&\qquad \sum_{p=1\dots 3} \left[ \chi_\perp^{\sf Q}(\partial_\tau\phi_p)^2 + \rho_{\sf d}^{\sf Q} \sum_{\lambda ={\sf x,y}}(\partial_\lambda\phi_p)^2 \right]\nonumber \\
& \qquad +\sum_{p=4\dots 6} \left[ \chi_\perp^{\sf S}(\partial_\tau\phi_p)^2 + \rho_{\sf d}^{\sf S} \sum_{\lambda ={\sf x,y}}(\partial_\lambda\phi_p)^2 +\chi_\perp^{\sf S} \Delta^2 \phi_p^2 \right]. \nonumber
\end{align}
The long-wavelength properties of the 3-sublattice AFQ state are completely characterised
by the four parameters $\chi^{\sf Q}_\perp$, $\chi_\perp^{\sf S}$, $\rho_{\sf d}^{\sf Q}=\rho_{\sf d}^{\sf S}$ and $\Delta$.
Table~\ref{table:3-sublattice-dictionary} in Section~\ref{sec:SU2} provides a ``dictionary'' for 
converting between the parameters of the continuum theory, and the parameters of the minimal 
microscopic model ${\mathcal H}^{\sf BBQ}_\triangle$~[Eq.~(\ref{eq:H-BBQ-triangle})].


The dispersion of the spin excitations of this spin-nematic state then follow from the usual 
Euler-Lagrange equations.
The three fields $\phi_1$, $\phi_2$ and $\phi_3$ describe 
Goldstone modes with linear dispersion $\omega_{\bf q}^{\sf Q}$ [Eq.~(\ref{eq:omega-})], viz~:
\begin{align}
\omega_{\bf q}^{\sf Q} \approx v_{\sf Q} |{\bf q}|, 
   \qquad v_{\sf Q}=\sqrt{\frac{\rho_{\sf d}^{\sf Q}}{\chi^{\sf Q}_\perp}} \nonumber
\end{align}
while the three fields $\phi_4$, $\phi_5$ and $\phi_6$, describe gapped
excitations with dispersion $\omega_{\bf q}^{\sf S}$ [Eq.~(\ref{eq:omega+})], viz~:
\begin{align}
\omega_{\bf q}^{\sf S} \approx \sqrt{\Delta^2+v_{\sf S}^2{\bf q}^2}, 
   \qquad v_{\sf S}=\sqrt{\frac{\rho_{\sf d}^{\sf S}}{\chi^{\sf S}_\perp}} \nonumber
\end{align}


The remaining challenge is to correctly reference the continuum theory back to
the lattice, and to calculate the intensities associated with each branch 
of excitation.
To do this it is necessary to transcribe the spin degrees of freedom $(S^x,S^y,S^z)$ 
in terms of the fields $\boldsymbol{\phi}$, and then decompose spin-spin 
correlations $\langle S^\alpha S^\beta \rangle$ as contractions of the 
$\boldsymbol{\phi}$ fields.
These can contain contributions from more than one kind of 
excitation. 
A worked example of this type of calculation is given in Appendix~A of Ref.~\protect\onlinecite{smerald11}.    


It follows from Eq.~(\ref{eq:isotropicspins}) [Section~\ref{sec:3-sublattice-spin-toolbox}] 
that, to leading order in $\phi$, 
\begin{align}
&\delta {\bf S}({\bf r},t) \approx  \nonumber \\
&-\frac{ \chi_\perp^{\sf Q} }{2} \left(
\begin{array}{c}
(2-e^{i{\bf k}_{\sf K}.{\bf r}}-e^{-i{\bf k}_{\sf K}.{\bf r}})   \partial_t  \phi_3  \\
(2 + [1+e^{-i\frac{2\pi}{3}}] e^{i{\bf k}_{\sf K}.{\bf r}} + [1+e^{i\frac{2\pi}{3}}] e^{-i{\bf k}_{\sf K}.{\bf r}})  \partial_t \phi_2  \\
(2 + [1+e^{i\frac{2\pi}{3}}] e^{i{\bf k}_{\sf K}.{\bf r}} + [1+e^{-i\frac{2\pi}{3}}] e^{-i{\bf k}_{\sf K}.{\bf r}})  \partial_t\phi_1
\end{array}
\right)  \nonumber \\
&+ \frac{2}{3} \left(
\begin{array}{c}
(e^{i\frac{2\pi}{3}}-e^{-i\frac{2\pi}{3}})(e^{i{\bf k}_{\sf K}.{\bf r}} -e^{-i{\bf k}_{\sf K}.{\bf r}}) \ \phi_6    \\
([-1+e^{-i\frac{2\pi}{3}}] e^{i{\bf k}_{\sf K}.{\bf r}} + [-1+e^{i\frac{2\pi}{3}}] e^{-i{\bf k}_{\sf K}.{\bf r}}) \ \phi_5  \\
([1-e^{i\frac{2\pi}{3}}] e^{i{\bf k}_{\sf K}.{\bf r}} + [1-e^{-i\frac{2\pi}{3}}] e^{-i{\bf k}_{\sf K}.{\bf r}}) \ \phi_4
\end{array}
\right) .
\label{eq:spinfield}
\end{align}
%
%
%
From Eq.~(\ref{eq:spinfield}), we can immediately identify the three fields 
$\phi_1$, $\phi_2$ and $\phi_3$ with quadrupole waves whose 
contribution to scattering vanishes as $\chi_\perp^{\sf Q} \partial_t \phi 
\sim \chi_\perp^{\sf Q} \omega_{\bf q}^{\sf Q}$.
Meanwhile, the three fields $\phi_4$, $\phi_5$ and $\phi_6$ are spin waves 
with a robust dipole moment.


\begin{figure*}[ht]
\includegraphics[width=0.7\textwidth]{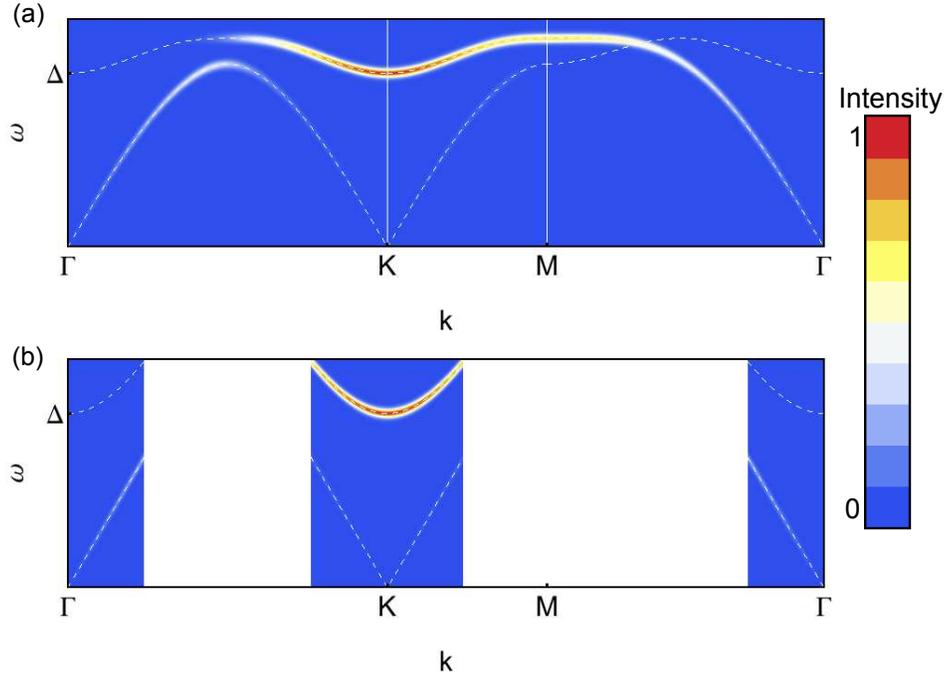}
\caption{\footnotesize{(Color online). 
Prediction for 
inelastic neutron scattering from a state with 3-sublattice antiferroquadrupolar (AFQ) 
spin nematic order of the type 
shown in Fig.~\ref{fig:AFNtriangle}.
Animations showing the nature of the spin-dipole fluctuations associated with the gapless\cite{AFQ-Qwave-animation} and gapped\cite{AFQ-Swave-animation} excitations are shown in the supplemental material.
(a) prediction of the microscopic 
`flavour wave' theory, as calculated from $\mathcal{H}_{\sf BBQ}$~[Eq.~(\ref{eq:H-BBQ-triangle})] 
for $J_1=1$, $J_2=1.22$ (cf. Ref.~[\onlinecite{tsunetsugu06,tsunetsugu07}]).  
The dashed white lines show the one-particle dispersion relations
$\omega_{\bf k}^{\pm}$~[Eq.~(\ref{eq:3subAFQflavourwave})], where the gap to spin-wave excitations is $\Delta = 6\sqrt{J_2(J_2-J_1)}$.
(b) prediction of the continuum theory 
$\mathcal{S}^{\sf SU(2)}_\triangle[\boldsymbol{\phi}]$~[Eq.~(\ref{eq:linaction})]
for the same set of parameters.
The dashed white lines show the one-particle dispersion relations $\omega_{\bf k}^{\sf Q}$~[Eq.~(\ref{eq:omega-})] and 
$\omega_{\bf k}^{\sf S}$~[Eq.~(\ref{eq:omega+})].
The majority of the spectral weight is found in the spin-wave band, in 
the vicinity of the 3-sublattice AFQ ordering vector 
${\bf k}_{\sf K} = (4\pi/3,0)$. 
All predictions have been convoluted with a gaussian of FWHM 
$0.042\Delta$ to mimic experimental resolution.
The circuit $\Gamma$-${\sf K}$-$M$-$\Gamma$ in reciprocal space 
is shown in Fig.~\ref{fig:mbz}. 
}}
\label{fig:3SLAFQ-chiS}
\end{figure*}


\subsubsection{Single particle scattering near to ${\bf k} = {\bf k}_{\sf K}$}


Let us consider first scattering involving a single excitation near to the ordering vector, ${\bf k} = {\bf k}_{\sf K}$.  
Here the field theory predicts a gapless Goldstone mode with dispersion 
$\omega_{\bf q}^{\sf Q}$ [Eq.~(\ref{eq:omega-})], for small ${\bf q}={\bf k}-{\bf k}_{\sf K}$. 
This is accompanied by a gapped 
spin-wave excitation with dispersion $\omega_{\bf q}^{\sf S}$ [Eq.~(\ref{eq:omega+})].
The associated single-particle contribution to the dynamical susceptibility is,
\begin{align}
\Im m \{\chi_{\sf S}^{\sf xx}({\bf k}_{\sf K}+{\bf q}, \omega)\} 
& \approx \frac{\pi}{8} (g\mu_{\sf B})^2 
\chi_\perp^{\sf Q} \omega_{\bf q}^{\sf Q}  
\delta(\omega -\omega_{\bf q}^{\sf Q}) \nonumber \\
&+\frac{2\pi}{3}(g\mu_{\sf B})^2 \frac{1}{\chi_\perp^{\sf S} \omega_{\bf q}^{\sf S}} 
\delta(\omega -\omega_{\bf q}^{\sf S}),
\label{eq:1pK}
\end{align}
where ${\bf q} \approx 0$.  
Scattering close to the $K^\prime$ point is exactly equivalent.
From Eq.~(\ref{eq:1pK}) we see that the intensity of scattering from the quadrupole wave 
vanishes as $\chi_\perp^{\sf Q} \omega_{\bf q}^{\sf Q} \sim \chi_\perp^{\sf Q} v_{\sf Q} |{\bf q}|$
for ${\bf q} \to 0$.  
Meanwhile the scattering from the spin-wave excitation is enhanced as 
\mbox{$1/(\chi_\perp^{\sf S} \omega_{\bf q}^{\sf S}) \sim 1/(\Delta \chi_\perp^{\sf S})$} in the same limit.
The spin-wave excitation will therefore dominate the response seen in experiment.
These features are illustrated in Fig.~\ref{fig:3SLAFQ-chiS}.


Exactly the same quadrupole and spin wave excitations are found in flavour-wave 
calculations~\cite{tsunetsugu06,tsunetsugu07} for the 3-sublattice AFQ phase of the spin-1 
BBQ model on the triangular lattice $\mathcal{H}_{\sf BBQ}$~[Eq.~(\ref{eq:H-BBQ-triangle})].
These predict a 1-particle contribution to the dynamical susceptibility which behaves as,
 \begin{align}
&\Im m \{\chi_{\sf S}^{\sf xx}({\bf k}, \omega)\} \approx \nonumber \\
& \qquad \pi (1+\cos\theta_{\bf k}) (g\mu_{\sf B})^2 \frac{J_2(1-|\gamma_{\bf k}|) }{ \omega^-_{\bf k}} 
\delta(\omega -\omega_{\bf k}^-) \nonumber \\
& \qquad +\pi (1-\cos\theta_{\bf k}) (g\mu_{\sf B})^2 \frac{J_2(1+|\gamma_{\bf k}|) }{ \omega^+_{\bf k}} 
\delta(\omega -\omega_{\bf k}^+),
\end{align}
where $\omega_{\bf k}^\pm$ is given in Eq.~(\ref{eq:3subAFQflavourwave}), 
$\gamma_{\bf k}$ in Eq.~(\ref{eq:gamma}) and,
\begin{align}
e^{i\theta_{\bf q}} = \frac{\gamma_{\bf k}}{|\gamma_{\bf k}|}.
\end{align}
Matching this to the predictions of the field theory for  ${\bf k} \approx {\bf k}_{\sf K}$, 
and translating parameters using Table~\ref{table:3-sublattice-dictionary}, we find exact, quantitative, 
agreement between the two approaches at long wavelength.   
This comparison is illustrated in Fig.~\ref{fig:3SLAFQ-chiS}.


\subsubsection{Single particle scattering near ${\bf k} = 0$}


Close to the $\Gamma$ point, we find a one-particle contribution to the dynamical susceptibility
\begin{align}
\Im m \{\chi_{\sf S}^{\sf xx}({\bf q}, \omega)\} 
 \approx \frac{\pi}{2} (g\mu_{\sf B})^2 \chi_\perp^{\sf Q}
\omega_{\bf q}^{\sf Q}  \delta(\omega -\omega_{\bf q}^{\sf Q}),
\label{eq:1pGamma}
\end{align}
where $\omega_{\bf q}^{\sf Q}$ is given by Eq.~(\ref{eq:omega-}) and ${\bf q}\approx 0$.
This corresponds to a linearly-dispersing quadrupole wave, whose intensity vanishes linearly 
for ${\sf q} \to 0$, but with $4$ times the prefactor for scattering near ${\bf k}_{\sf K}$.   
Once again this result  is in quantitative agreement with the predictions of the lattice model 
$\mathcal{H}_{\sf BBQ}$~[Eq.~(\ref{eq:H-BBQ-triangle})]. 


While spin-wave excitations are defined for all ${\bf k}$, they do not contribute 
to single-particle scattering in the vicinity of the $\Gamma$ point.  
This is because the dipole fluctuations on neighbouring sublattices are exactly in 
anti-phase, and therefore cancel for ${\bf k} \to 0$.   
This cancellation is not accidental, but required by the ${\sf SU(2)}$ symmetry of the spin-nematic state, and is a manifestation of the sum rule, Eq.~(\ref{eq:sumrule}).
%


\subsubsection{Adding it all up}


Fig.~\ref{fig:3SLAFQ-chiS} shows the result of summing all the 1-particle 
contributions to the $T=0$ dynamic susceptibility, to give an overall prediction for inelastic 
neutron scattering from a 3-sublattice AFQ state.
Most of the spectral weight resides close to the ${\sf K}$- and ${\sf K}'$-points, 
in the spin-wave band. 
In contrast, quadrupole-waves contribute very little to scattering.


\section{Dynamical quadrupolar susceptibility}
\label{sec:quad-suscep}


It is possible that a resonant technique, such as resonant x-ray scattering, could directly probe the 4-spin correlation function.
This would provide access to the dynamical quadrupolar susceptibility.
In the f-electron system UPd$_3$, resonant x-ray scattering has been used to access the quadrupolar order parameter\cite{mcmorrow01,walker06,walker08}.
While this is a different type of quadrupolar order, formed from a combination of spin and orbital degrees of freedom, the idea may carry over to the pure spin quadrupole considered in this publication.
We therefore present predictions for $\Im m \{\chi_{\sf Q}^{\alpha\beta\gamma\delta}({\bf k}, \omega)\}$.


\subsection{Quadrupolar excitations in a 3-sublattice AFQ state}


It follows from Eq.~(\ref{eq:Qmoments}) [Section~\ref{sec:3-sublattice-spin-toolbox}] 
that, to linear order in $\phi$, 
\begin{align}
&\delta {\bf Q}({\bf r},t) \approx  \nonumber \\
&- \frac{2}{3} \left(
\begin{array}{c}
0 \\
0 \\
([1-e^{i\frac{2\pi}{3}}] e^{i{\bf k}_{\sf K}.{\bf r}} + [1-e^{-i\frac{2\pi}{3}}] e^{-i{\bf k}_{\sf K}.{\bf r}}) \ \phi_1 \\
(e^{i\frac{2\pi}{3}}-e^{-i\frac{2\pi}{3}})(e^{i{\bf k}_{\sf K}.{\bf r}} -e^{-i{\bf k}_{\sf K}.{\bf r}}) \ \phi_3  \\
([1-e^{-i\frac{2\pi}{3}}] e^{i{\bf k}_{\sf K}.{\bf r}} + [1-e^{i\frac{2\pi}{3}}] e^{-i{\bf k}_{\sf K}.{\bf r}}) \ \phi_2
\end{array}
\right) \nonumber \\
&-\frac{ \chi_\perp^{\sf S} }{2} \hspace{-1mm} \left( \hspace{-1mm}
\begin{array}{c}
0 \\
0 \\
(2 + [1+e^{i\frac{2\pi}{3}}] e^{i{\bf k}_{\sf K}.{\bf r}} + [1+e^{-i\frac{2\pi}{3}}] e^{-i{\bf k}_{\sf K}.{\bf r}})  \partial_t\phi_4 \\
(2-e^{i{\bf k}_{\sf K}.{\bf r}}-e^{-i{\bf k}_{\sf K}.{\bf r}})   \partial_t  \phi_6  \\
-(2 + [1+e^{-i\frac{2\pi}{3}}] e^{i{\bf k}_{\sf K}.{\bf r}} + [1+e^{i\frac{2\pi}{3}}] e^{-i{\bf k}_{\sf K}.{\bf r}})  \partial_t \phi_5  \\
\end{array}
\hspace{-1mm} \right)  
\label{eq:Qfield}
\end{align}
In this basis, and at leading order in the perturbation expansion, the only non-zero entries in the susceptibility tensor are,
\begin{align}
\Im m \{\chi_{\sf Q}^{\sf xyxy}({\bf k}, \omega)\} = 
\Im m \{\chi_{\sf Q}^{\sf yzyz}({\bf k}, \omega)\} =
\Im m \{\chi_{\sf Q}^{\sf xzxz}({\bf k}, \omega)\} ,
\end{align}
and those related by the symmetry of the ${Q^{\alpha \beta}}$ tensor.


From Eq.~(\ref{eq:Qfield}) we can see that the $\phi_1$, $\phi_2$ and $\phi_3$ Goldstone mode fields give a diverging contribution to the quadrupolar susceptibility approaching the Bragg peak at ${\bf k}={\bf k}_{\sf K}$.
Conversely, the quadrupole fluctuations induced dynamically by the gapped, spin wave modes are small as \mbox{$\chi_\perp^{\sf S} \partial_t \phi \sim \omega_{\bf q}^{\sf S}$}.


\begin{figure*}[ht]
\includegraphics[width=0.7\textwidth]{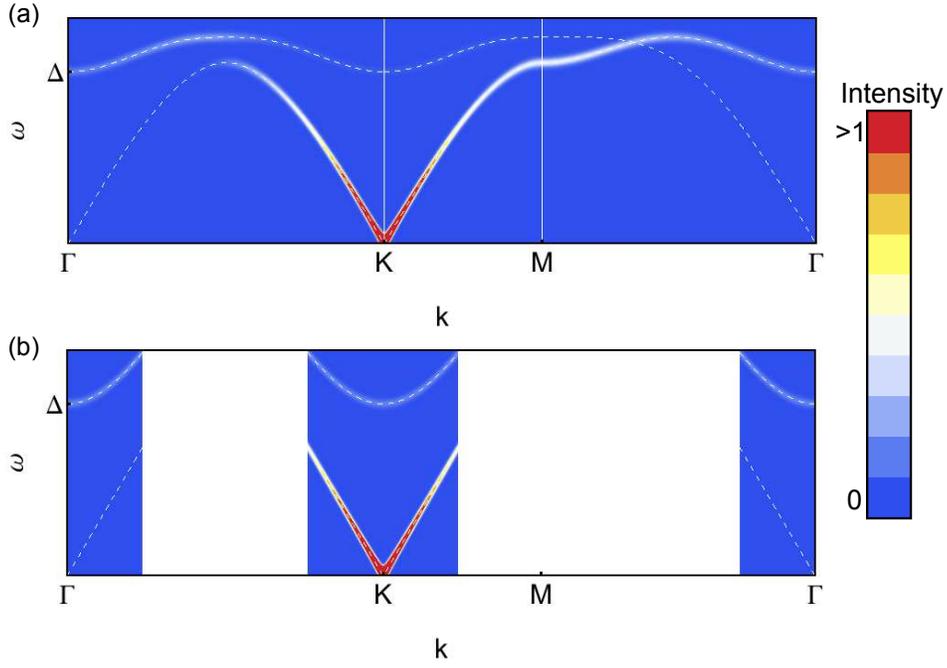}
\caption{\footnotesize{(Color online). 
Prediction for 
the dynamical quadrupolar susceptibility, $\Im m \{\chi_{\sf Q}^{\sf xyxy}({\bf k}, \omega)\}$, for a state with 3-sublattice antiferroquadrupolar (AFQ) 
spin nematic order of the type 
shown in Fig.~\ref{fig:AFNtriangle}.
Animations showing the nature of the spin-dipole fluctuations associated with the gapless\cite{AFQ-Qwave-animation} and gapped\cite{AFQ-Swave-animation} excitations are shown in the supplemental material.
(a) prediction of the microscopic 
`flavour wave' theory, as calculated from $\mathcal{H}_{\sf BBQ}$~[Eq.~(\ref{eq:H-BBQ-triangle})] 
for $J_1=1$, $J_2=1.22$.  
The dashed white lines show the one-particle dispersion relations
$\omega_{\bf k}^{\pm}$~[Eq.~(\ref{eq:3subAFQflavourwave})], where the gap to spin-wave excitations is $\Delta = 6\sqrt{J_2(J_2-J_1)}$.
(b) prediction of the continuum theory 
$\mathcal{S}^{\sf SU(2)}_\triangle[\boldsymbol{\phi}]$~[Eq.~(\ref{eq:linaction})]
for the same set of parameters.
The dashed white lines show the one-particle dispersion relations $\omega_{\bf k}^{\sf Q}$~[Eq.~(\ref{eq:omega-})] and 
$\omega_{\bf k}^{\sf S}$~[Eq.~(\ref{eq:omega+})].
The dominant feature is a diverging `Bragg peak' associated with the gapless, quadrupole-wave mode in 
the vicinity of the 3-sublattice AFQ ordering vector 
${\bf k}_{\sf K} = (4\pi/3,0)$. 
All predictions have been convoluted with a gaussian of FWHM 
$0.042\Delta$ to mimic experimental resolution.
The circuit $\Gamma$-${\sf K}$-$M$-$\Gamma$ in reciprocal space 
is shown in Fig.~\ref{fig:mbz}. 
}}
\label{fig:3SLAFQ-chiQ}
\end{figure*}


\subsubsection{Single particle scattering near to ${\bf k} = {\bf k}_{\sf K}$}


The dynamical quadrupolar susceptibility can be determined in an analogous manner to the spin susceptibility [see Section~\ref{sec:chiS}].
Close to ${\bf k} = {\bf k}_{\sf K}$ the field theory predicts,
\begin{align}
\Im m \{\chi_{\sf Q}^{\sf xyxy}({\bf k}_{\sf K}+{\bf q}, \omega)\} 
& \approx \frac{2\pi}{3}(g\mu_{\sf B})^4 \frac{1}{\chi_\perp^{\sf Q} \omega_{\bf q}^{\sf Q}} 
\delta(\omega -\omega_{\bf q}^{\sf Q}) \nonumber \\
&+\frac{\pi}{8} (g\mu_{\sf B})^4 
\chi_\perp^{\sf S} \omega_{\bf q}^{\sf S}  
\delta(\omega -\omega_{\bf q}^{\sf S}) ,
\label{eq:chiQ-Kpoint}
\end{align}
where ${\bf q} \approx 0$.  
Scattering close to the $K^\prime$ point is exactly equivalent.
Eq.~(\ref{eq:chiQ-Kpoint}) shows that the intensity of scattering due to the quadrupolar modes diverges as \mbox{$1/(\chi_\perp^{\sf Q}\omega_{\bf q}^{\sf Q}) \sim 1/|{\bf q}|$} for ${\bf q}\to 0$.
Thus there is a `Bragg peak' in the quadrupolar susceptibililty, as one expects for quadrupolar order.
The gapped spin-wave modes induce a small quadrupole fluctuation, and this gives only a weak contribution to the susceptibility.


Linear flavour wave theory for the spin-1 BBQ model on the triangular lattice, $\mathcal{H}_{\sf BBQ}$~[Eq.~(\ref{eq:H-BBQ-triangle})], predicts,
  \begin{align}
&\Im m \{\chi_{\sf Q}^{\sf xyxy}({\bf k}, \omega)\} \approx \nonumber \\
& \pi (1-\cos\theta_{\bf k}) (g\mu_{\sf B})^4 
\frac{J_2(1-|\gamma_{\bf k}|) +2J_1|\gamma_{\bf k}|}{ \omega^-_{\bf k}} 
\delta(\omega -\omega_{\bf k}^-) \nonumber \\
& +\pi (1+\cos\theta_{\bf k}) (g\mu_{\sf B})^4 
\frac{J_2(1+|\gamma_{\bf k}|) -2J_1|\gamma_{\bf k}|}{ \omega^+_{\bf k}} 
\delta(\omega -\omega_{\bf k}^+),
\label{eq:chiQ-flvwv}
\end{align}
and this is quantitative agreement with the field theory, Eq.~(\ref{eq:chiQ-Kpoint}), approaching the high symmetry points.


\subsubsection{Single particle scattering near ${\bf k} = 0$}


Close to the $\Gamma$ point, we find a one-particle contribution to the dynamical quadrupolar susceptibility,
\begin{align}
\Im m \{\chi_{\sf Q}^{\sf xyxy}({\bf q}, \omega)\} 
 \approx \frac{\pi}{2} (g\mu_{\sf B})^4 \chi_\perp^{\sf S}
\omega_{\bf q}^{\sf S}  \delta(\omega -\omega_{\bf q}^{\sf S}).
\label{eq:chiQ-Gamma}
\end{align}
The quadrupole fluctuations induced dynamically by the gapped, spin-wave modes are suppressed by a factor $\chi_\perp^{\sf S}\omega_{\bf q}^{\sf S}$ and have low intensity compared to the diverging Goldstone mode at the {\sf K} point.
One interesting feature is that the gapless quadrupole mode at the $\Gamma$ point does not appear in the field theory calculation of the susceptibility, due to the fact that neighbouring quadrupoles beat in antiphase [see Eq.~(\ref{eq:Qmoments})]. 
This is in agreement with the flavour wave theory, Eq.~(\ref{eq:chiQ-flvwv}), where the susceptibility turns on very slowly as $\Im m \{\chi_{\sf Q}^{\sf xyxy}({\bf q}, \omega)\}\sim q^5 $.


\subsubsection{Adding it all up}


Fig.~\ref{fig:3SLAFQ-chiQ} shows the result of summing all the 1-particle 
contributions to the $T=0$ dynamic quadrupolar susceptibility, to give an overall prediction for scattering from a 3-sublattice AFQ state.
The dominant feature is the presence of `Bragg peaks' at the ${\sf K}$- and ${\sf K}'$-points.
There is also a faint band where the gapped, spin-wave excitations dynamically induce a small, fluctuating quadrupole moment.


\section{Discussion and Conclusions}
\label{sec:conclusions}


Spin-nematic order remains an enigma.
First proposed almost 40 years ago, and now studied in a wide range of 
theoretical models, it has never yet been unambiguously observed in experiment.
Much of the difficulty in identifying a spin-nematic state arises from
the fact that spin-nematic order does not break time-reversal symmetry.
As a consequence, it cannot give rise to the internal magnetic
fields measured by the common probes of static magnetic 
order --- neutron scattering, NMR and muon spin rotation.
In this respect, spin-nematic order has much in common with  
multipolar `hidden order' phases in rare earth magnets\cite{santini09}.
In principle, spin nematic order {\it could} be probed through its 
excitations.
However, because of the complexity of the problem, 
these remain relatively poorly understood.


In this paper we have attempted to narrow the gap between theory 
and experiment, by constructing a continuum field theory of a
three-sublattice antiferroquadrupolar (AFQ) spin-nematic state.
This field theory offers a `model-independent' approach to interpreting 
experiment, and can be used to explore the physical nature of the magnetic 
excitations of AFQ states.
In the absence of magnetic field, we find that the long-wavelength excitations 
of AFQ states naturally divide into a set of three gapless, 
quadrupole-wave modes, --- the Goldstone modes --- 
together with three gapped excitations with a strong spin-dipole character.


This field theory can also be used to make concrete predictions for the 
fluctuating spin-dipole fields associated with each type of excitation, and its
associated signature in experiment.
In this paper we have focused on the most direct probe of spin-dipole 
fluctuations --- inelastic neutron scattering.
We find that quadrupole waves couple only weakly with neutrons, with 
the intensity of scattering vanishing linearly at low energies.
However the gapped modes possess a substantial dipole
moment and couple strongly to neutrons.
The observation of this gapped excitation, together with a set of ghostly low-energy 
Goldstone modes, in the {\it absence} of magnetic Bragg peaks, would 
constitute strong evidence for AFQ spin-nematic order.


Finally we make predictions for the dynamical quadrupole susceptibility.
This exhibits diverging Bragg-peak like intensity approaching the Goldstone modes, 
along with a very faint gapped mode.
As in the f-electron systems, this may be measurable using resonant x-ray scattering.
Such experiments would directly probe the order parameter, and could in consequence provide compelling 
evidence for the existence of spin-nematic order.
How these excitations evolve with field, and what their consequences are for 
NMR $1/T_1$ relaxation rates will be explored in separate 
publications\cite{smerald-unpub,smerald11,smerald-arXiv}.


%
An obvious question for future work is the role of interactions.
As in the case of FQ order [\onlinecite{baryakhtar13}], interactions between the Goldstone modes 
of the AFQ state endow these excitations with a finite, $k^2$-dependent lifetime.
There is also a corresponding renormalisation of the director stiffness, $\rho_{\sf d}^{\sf Q}$, 
leading to small changes in the velocity of the Goldstone modes.
However, the most interesting features come from the interaction between the 
Goldstone modes and the gapped, long-wavelength ``spin-wave'' modes.
This is true both from an experimental point of view, since the gapped
modes support large spin-dipole fluctuations, and a theoretical point of view,
where these type of interactions have not been as thoroughly explored
as those between Goldstone modes.
We will return to these effects in a future paper\cite{smerald-unpub}.


In conclusion, the ${\sf SU(3)}$ generalisation of the non-linear sigma model developed
in this text provides a robust means of characterising spin-nematic states with 
antiferroquadrupolar order, which is independent of any particular microscopic model.
This sigma model approach provides an excellent starting point for understanding
the universal behaviour of spin-nematic states, and leads to concrete, testable 
predictions for experiment.
For this reason, it can serve as an important tool for establishing whether spin-nematic 
order exists in a wide variety of real materials.
We hope that the waves predicted by the sigma model will, in the near future, be seen.


{\it Acknowledgments.}   
We are grateful to Tsutomu Momoi for a number helpful comments on this work, and 
to Karlo Penc for a careful reading of the manuscript.
This research was supported under EPSRC grants EP/C539974/1 and EP/G031460/1

%
%



\begin{thebibliography}{99}
\bibitem{AFQ-Swave-animation} 
Animation showing the nature of the spin-wave modes, calculated directly from the 
linearised continuum field theory.
%
A large, coherent fluctuation in the spin-dipole field, is accompanied by a small, 
dynamically generated quadrupole fluctuation.
%
The animation shows an equally weighted superposition of the fields 
$\phi_4$, $\phi_5$ and $\phi_6$.  
%
(See [\onlinecite{AFQ-Qwave-animation}] for further details.)
\end{thebibliography}
\end{document}